\documentclass[fleqn,usenatbib]{mnras}

\usepackage{newtxtext,newtxmath}
\usepackage[T1]{fontenc}

\DeclareRobustCommand{\VAN}[3]{#2}
\let\VANthebibliography\thebibliography
\def\thebibliography{\DeclareRobustCommand{\VAN}[3]{##3}\VANthebibliography}

\usepackage{graphicx}	% Including figure files
\usepackage{amsmath}	% Advanced maths commands
\usepackage{gensymb}
\usepackage{orcidlink}
\usepackage{color} %Just for markup while writing
\usepackage{hyperref}
\usepackage{cleveref}
\usepackage{verbatim}

\newcommand{\kms}{\mathrm{km\,s^{-1}}}
\newcommand{\Msun}{\mathrm{M_\odot}}

\defcitealias{Fielding2022}{FB22}
\defcitealias{Smith2024}{Paper I}
\defcitealias{Springel2003}{SH03}

\crefname{section}{Section}{Sections}
\Crefname{section}{Section}{Sections}
\crefname{subsection}{Section}{Sections}
\Crefname{subsection}{Section}{Sections}
\crefname{subsubsection}{Section}{Sections}
\Crefname{subsubsection}{Section}{Sections}
\crefname{figure}{Fig.}{Fig.}
\Crefname{figure}{Fig.}{Fig.}
\crefname{equation}{equation}{equations}
\Crefname{equation}{Equation}{Equations}
\creflabelformat{equation}{#2#1#3}
\crefname{table}{Table}{Tables}
\Crefname{table}{Table}{Tables}
\crefname{appendix}{Appendix}{Appendices}
\Crefname{appendix}{Appendix}{Appendices}
\crefname{enumi}{point}{points}
\Crefname{enumi}{Point}{Points}
\title[Arkenstone II]{\textsc{Arkenstone} - II. A model for unresolved cool clouds entrained in galactic winds in cosmological simulations}

\author[M. C. Smith et al.]{
Matthew C. Smith\orcidlink{0000-0002-9849-877X},$^{1}$\thanks{E-mail: msmith@mpa-garching.mpg.de}
Drummond B. Fielding\orcidlink{0000-0003-3806-8548},$^{2,3}$
Greg L. Bryan\orcidlink{0000-0003-2630-9228},$^{4}$
Jake S. Bennett\orcidlink{0000-0002-8573-2993},$^{5}$
\newauthor
\ Chang-Goo Kim\orcidlink{0000-0003-2896-3725},$^{6}$
Eve C. Ostriker
\orcidlink{0000-0002-0509-9113}$^{6}$
and
Rachel S. Somerville\orcidlink{0000-0002-6748-6821}$^2$
\\
$^{1}$Max-Planck-Institut f{\"u}r Astrophysik, Karl-Schwarzschild-Str. 1, D-85748, Garching, Germany\\
$^{2}$Center for Computational Astrophysics, Flatiron Institute, 162 5\textsuperscript{th} Avenue, New York, NY 10010, USA\\
$^{3}$Department of Astronomy, Cornell University, Ithaca, NY 14853, USA\\
$^{4}$Department of Astronomy, Columbia University, 550 West 120\textsuperscript{th} Street, New York, NY 10027, USA\\
$^{5}$Center for Astrophysics | Harvard \& Smithsonian, 60 Garden Street, Cambridge, MA 02138, USA\\
$^{6}$Department of Astrophysical Sciences, Princeton University, 4 Ivy Lane, Princeton, NJ 08544, USA\\
}

\date{Accepted XXX. Received YYY; in original form ZZZ}

\pubyear{2024}

\begin{document}
\label{firstpage}
\pagerange{\pageref{firstpage}--\pageref{lastpage}}
\maketitle

\begin{abstract}
\textsc{Arkenstone} is a new scheme that allows multiphase, stellar feedback-driven winds to be included in coarse resolution cosmological simulations.
The evolution of galactic winds and their subsequent impact on the circumgalactic medium are altered by
exchanges of mass, energy, momentum, and metals between their component
phases. These exchanges are governed by complex, small-scale
physical processes that cannot be resolved in cosmological simulations.
In this second presentation paper, we describe \textsc{Arkenstone}'s
novel cloud particle approach for modelling unresolvable cool clouds entrained in
hot, fast winds. This general framework allows models of
the cloud--wind interaction, derived from
state-of-the-art high-resolution
simulations, to be applied in a large-scale context.
In this work, we adopt a cloud evolution model that captures
simultaneous cloud mass loss to and gain from
the ambient hot phase via turbulent mixing and 
radiative cooling, respectively.
We demonstrate the scheme using non-cosmological idealized
simulations of a galaxy with a realistic
circumgalactic medium component, using the \textsc{Arepo} code.
We show that the ability of a high-specific energy wind component
to perform preventative feedback may be limited by heavy loading of cool clouds coupled into it.
We demonstrate that the diverging evolution of
clouds of initially differing masses leads to a complex velocity field for the cool phase and a cloud mass function
that varies both spatially and temporally in a non-trivial manner.
These latter two phenomena can manifest in the simulation because of
our choice of a Lagrangian
discretisation of the cloud population, in contrast to other
proposed schemes. This is a Learning the Universe publication.
\end{abstract}


%%%%%%%%%%%%%%%%%%%%%%%%%%%%%%%%%%%%%%%%%%%%%%%%%%

%%%%%%%%%%%%%%%%% BODY OF PAPER %%%%%%%%%%%%%%%%%%

\section{Introduction} \label{sec:intro}
Galactic winds are observed to be a fundamental component of the
cosmic baryon cycle \citep{Tumlinson2017}
and therefore play an essential role in galaxy formation theory \citep{Somerville2015,Naab2017}.
As gas inflows transport mass from the intergalactic medium (IGM),
through the circumgalactic medium (CGM) to the interstellar medium (ISM)
of galaxies
\citep{Keres2005,Dekel2006},
large-scale winds driven by feedback from stars or active galactic
nuclei (AGN) provide a counterbalance, transporting
mass and energy outwards.
They can therefore regulate galactic star formation
as an ejective feedback process, removing gas from
the ISM before it can be converted into stars
\citep[e.g.][]{Mathews1971,Larson1974,Chevalier1985,Dekel1986,White1991},
or preventative, reducing inflows from the CGM/IGM
\citep[e.g.][]{Oppenheimer2010,vandeVoort2011,Dave2012,Lu2015,Lu2017,Pandya2020,Carr2023}.
They also shape the properties of the CGM,
influencing its baryon fraction, phase structure, and metallicity.

Galactic winds are often parameterised
in terms of loading factors.
The mass loading factor of a stellar
feedback driven wind is the ratio of
the emergent mass outflow rate to the
star formation rate (SFR) of the galaxy
driving it.
Likewise, the energy loading
is the emergent power of the wind
relative to the power generated by
stellar feedback (i.e. an efficiency factor, generally relative to the supernova input).
The specific energy of the wind (i.e. its
temperature and velocity) depends on the ratio
of the energy and mass loading factors.
The impact of a wind on its host CGM 
and the manner in which it regulates star formation
depends not only
on the mass outflow rate but on its specific energy \citep{Carr2023,Voit2024a,Voit2024b}.
Two winds can have the same energy loading
but different combinations of mass loading and specific energy.
A highly mass loaded wind is a source of ejective feedback,
removing mass from the ISM. However, for a fixed energy loading,
increasing the mass loading reduces the specific energy of the wind;
the available energy is diluted by being coupled to increasing
amounts of mass.
Such a wind has limited ability to impact the CGM, tending to
form fountain flows that return to the galaxy.
On the other hand, a high specific energy wind (with a low mass loading factor)
can transmit energy to large volumes of the CGM and beyond,
heating it and suppressing inflows (i.e. preventative feedback), but has little impact on the gas which
is already in the ISM.

This picture is complicated by the fact
that characterising a wind with a single
mass outflow rate and specific energy
is a substantial oversimplification.
In fact, they are observed
to be highly multiphase with detections of components
with temperatures of
$\lesssim 100$K \citep[e.g.,][]{Rupke2005,Bolatto2013,Martini2018},
$\sim 10^4$K \citep[e.g.,][]{Martin2009,Westmoquette2009,Nielsen2015},
$\sim 10^{5.5}$K \citep[e.g.,][]{Steidel2010,Kacprzak2015,Chisholm2018},
and $\gtrsim 10^7$K \citep[e.g.,][]{Strickland2009,Lopez2020,Hodges-Kluck2020}.
High resolution simulations that can resolve the generation
of multiphase outflows by stellar feedback show that
different components of the wind carry different
fractions of the total mass and energy loadings \citep[e.g.][]{Kim2018,Schneider2020,Rathjen2023,Steinwandel2024}.
For example, \cite{Fielding2018} and \cite{Kim2020b}
find that as the wind leaves the ISM a hot ($\gtrsim 10^6$K), fast component 
carries the majority of the energy while a cool ($\lesssim 10^4$K), slow component
carries most of the mass. With idealised simulations, \cite{Tan2024}
demonstrate the seeding of individual cool clouds into the hot wind
as the swept-up ISM fragments while the wind is generated by clustered supernovae (SNe).

A further complication is that as the wind
flows away from the galaxy, the various phases will interact,
redistributing mass, momentum and energy between them 
and impacting the large-scale evolution of the wind
(\citealt{Fielding2022}, hereafter \citetalias{Fielding2022}).
As demonstrated in ``cloud-crushing'' simulations \citep{Klein1994,Xu1995}
a cool cloud with radius $r_\mathrm{cl} $ located within a hot, fast moving
ambient wind
is accelerated by drag/ram pressure on a time-scale $t_\mathrm{drag} = \chi r_\mathrm{cl} / v_\mathrm{rel}$
but tends to be destroyed by Kelvin-Helmholtz and Rayleigh-Taylor instabilities on a cloud-crushing time-scale
$t_\mathrm{cc} = \chi^{1/2} r_\mathrm{cl} / v_\mathrm{rel}$, where $\chi$ and $v_\mathrm{rel}$
are the density contrast and relative velocity between the cloud and wind, respectively.
For clouds in hot galactic winds, $\chi \gg 1$, so the time-scale for cloud destruction
is shorter than that for cloud acceleration \citep{Zhang2017}. However, the picture is modified
by the inclusion of various additional physical processes such as
radiative cooling \citep[e.g.][]{Mellema2003,Cooper2009,Scannapieco2015},
thermal conduction \citep[e.g.][]{Marcolini2005,Orlando2005,Bruggen2016},
magnetic fields \citep[e.g.][]{MacLow1994,McCourt2015,Schneider2017,Cottle2020}
and cosmic rays \citep[e.g.][]{Wiener2019,Bruggen2020}.
Under particular circumstances,
radiative cooling can cause the cloud to accrete
mass from the hot phase \citep[e.g.][]{Marinacci2010,Armillotta2016,Gronke2018}.
Along the cloud--wind boundary, shearing motions
drive turbulence within a mixing layer between
the two phases. The mixing of the hot wind and
cool cloud material generates gas of intermediate
temperature ($\sim10^5\,\mathrm{K}$) which can
efficiently cool into the cloud.
Recent high resolution
studies of these turbulent radiative mixing
layers (TRMLs) have begun to constrain
the complex physics which governs the
rate at which hot wind material is accreted
into the cloud \citep[e.g.][]{Ji2019,Mandelker2020,Fielding2020,Tan2021b,Abruzzo2024}.
Importantly, this material brings momentum along
with its mass, providing a potentially significant
form of acceleration for the clouds. This was demonstrated
by \cite{Melso2019} in cloud inflow simulations,
by \cite{Vijayan2020} from analysis of outflows in simulations of star-forming disks,
and in \cite{Schneider2020} on kpc scales with 5 pc uniform
spatial resolution idealised simulations of a starbursting
galaxy.

Unfortunately, the resolution requirements for capturing
the generation and subsequent evolution of a multiphase wind
in an a priori manner in a simulation are very stringent.
When injecting energy from individual SNe directly into the
ISM, a mass resolution $\lesssim100\ \mathrm{M_\odot}$
is required just to get convergent bulk mass and energy
loadings \citep{Smith2018,Hu2019}, let alone obtain
the correct multiphase structure.
In order to approximately capture the
mass balance between the phases,
involving the interaction between
cool clouds and hot winds,
a mass resolution $\lesssim 1 \mathrm{M_\odot}$
is required \citep[e.g.][]{Gronke2020,Abruzzo2022,Gronke2022}.
This resolution is barely achievable in
cosmological ``zoom-in'' simulations of the lowest mass
dwarf galaxies.
Meanwhile, cosmological volume simulations,
necessary to build statistically significant
samples of galaxy evolution,
typically have mass resolutions of $\sim10^5 - 10^9\ \Msun$.
The highest resolution simulations in this class do
see multiphase structure in galactic outflows and the CGM in general \citep[e.g.][]{Nelson2019,Mitchell2020},
but these cannot reliably resolve the interactions
between the phases. The situation improves significantly if computing resources
are concentrated on resolving the CGM of
a single galaxy, producing a rich multiphase structure
of cool clouds suspended in a volume filling medium
\citep{vandeVoort2019,Peeples2019,Suresh2019,Hummels2019,Bennett2020,Ramesh2024b},
but the total mass of cool material in the CGM typically
continues to increase as the spatial resolution is improved.
These simulations do begin to allow individual, relatively massive cloud-like structures
to be identified and processes affecting their evolution
to be studied \cite[e.g.][]{Ramesh2024a}, but the spatial
resolution is still several orders of magnitude coarser
than what is required to properly capture the complex
mixing processes that should occur on their boundaries.
Regardless, the CGM zoom-in technique is computationally
intractable to apply to cosmological volumes.

Even assuming the most optimistic estimates of increases
in computing power and code efficiency,
we will not gain the ability to fully resolve
multiphase galactic wind and CGM material in cosmological
volume simulations for many years. 
An alternative approach is to
avoid attempting to resolve the multiphase material natively
and instead develop effective models that represent
the large-scale impact of the unresolvable
multiphase gas in a ``subgrid'' manner.
This is the approach we take in this work.
While the small-scale physics
must necessarily be modelled in a simplified manner,
this permits them to be included in a clean, interpretable
fashion. For many science questions, this is
preferable to poorly resolving or entirely omitting
the relevant processes.
\cite{Huang2020} presents a subgrid scheme for cool clouds
in galactic winds with a particle based approach
that models their disintegration, modulated by the effects
of thermal conduction. \cite{Weinberger2023} and \cite{Butsky2024}
present Eulerian multifluid approaches to treat
two phases of gas within the same resolution element. We
will discuss these three schemes in detail in \cref{subsec:comparison}.

\textsc{Arkenstone} is a new subgrid model for
stellar feedback-driven galactic winds
that emphasises their multiphase nature. The scheme is
intended for large volume cosmological simulations
where neither the multiphase structure of the ISM, galactic winds
or CGM can be resolved.
\textsc{Arkenstone} has been specifically designed to
provide a flexible framework for coarse-graining
results from analytical models and very high resolution
numerical studies of relevant small-scale physics.
\textsc{Arkenstone} is
developed in concert with the Learning the Universe (LtU)
Collaboration.\footnote{\url{http://learning-the-universe.org}}
LtU will deploy \textsc{Arkenstone} in the next generation
of cosmological volume simulations, alongside
improved models for the ISM and star formation.
While distinct from \textsc{Arkenstone}, these new ISM models
(Hassan et al. in press will present a prototype) will be
calibrated to ISM patch simulations \citep{Kim2024}
that can also be used to inform the input
parameters for \textsc{Arkenstone} \citep[see e.g.][]{Kim2020a}.

\textsc{Arkenstone} uses a wind particle propagation scheme
to inject winds immediately outside of the ISM, granting
very fine control over the properties of the wind at launch
irrespective of the coarse resolution that must be adopted
in this type of simulation. The model has three
novel features:
\begin{enumerate}
\item Winds are launched with hot and cool components with
separate mass and energy loadings, inspired by the results
from high resolution simulations, as mentioned above.

\item The hot, fast phase of the wind is injected and evolved
with a new ``displacement recoupling'' and refinement scheme
that properly treats high-specific energy, low density flows.

\item The cool phase is modelled using ``cloud particles''
to represent clouds embedded in the hot flow.
These particles exchange mass, energy, momentum,
and metals bidirectionally with the ambient hot wind.
\end{enumerate}

We first presented \textsc{Arkenstone} in
\cite{Smith2024} (hereafter \citetalias{Smith2024}),
detailing the first two aspects of the scheme. In particular,
we demonstrated that without the techniques employed by our scheme
it is impossible to resolve the high-specific energy winds
expected to drive preventative feedback processes.
This is because the low densities inherent to high-specific energy
winds cause poor spatial resolution when quasi-Lagrangian
refinement strategies (ubiquitous in cosmological simulations
carried out with both Lagrangian and Eulerian codes) are used.
In particular, failing to resolve the sonic point of a wind
results in an incorrect evolution of the balance of kinetic and thermal energy in the wind as it flows outwards. \textsc{Arkenstone}
makes it possible to properly consider preventative feedback
that originates from SNe
in cosmological simulations. We refer the reader to \citetalias{Smith2024}
for full details of the scheme, as well as a detailed discussion
of other stellar feedback and galactic wind schemes in the literature.

In this work, we present the remaining aspect of \textsc{Arkenstone}:
the cloud particle scheme. This is a framework
with which results drawn from
state-of-the-art cloud-crushing and TRML simulations,
that capture the fine details of the relevant small-scale physics,
can be included in large-scale cosmological simulations. The implementation
is agnostic to the choice of cloud--wind interaction model,
so long as fluxes of mass, momentum, energy and metals between
the phases can be predicted. For this first demonstration,
we use the model of \citetalias{Fielding2022}. In \cref{sec:methods}
we present our numerical methodology. In \cref{sec:results}
we apply our scheme to a series of
idealised simulations of isolated galaxies (carried out at a resolution
achievable in a cosmological volume simulation), 
demonstrating the general behaviour of the model and
highlighting various interesting regimes.
In \cref{sec:discussion} we discuss future applications of the
model, the interpretation of its predictions and possible
extensions to included physical processes. We also
compare our scheme to other relevant approaches in the literature.
We summarise our findings in \cref{sec:conclusion}.

\subsection{Nomenclature}
We are concerned with multiphase winds
comprised of various components, as described above.
However, much of the cloud-crushing literature refers
to cloud--wind interactions. Of course, once entrained,
one can consider cool material as being part of the multiphase
wind. For clarity, we adopt the following conventions.
``Hot wind'' is used as a short hand for high-specific energy
galactic winds (which are hot, fast and low density).
Aspects of the \textsc{Arkenstone} scheme relating to the
modelling of hot winds were described in detail 
in \citetalias{Smith2024} and are referred to as \textsc{Arkenstone-Hot}.
We generate the hot wind component
with the use of ``wind particles''
(a summary of the methodology can be found in \cref{subsec:launch_and_recouple}).
Cool clouds are modelled with ``cloud particles''.  Here, we use ``cool'' to refer to any gas with temperature $\leq 10^4\,\mathrm{K}$; while ``cold'' is often used for this thermal range in the field of galaxy formation, in the ISM literature the term ``cold'' is reserved for atomic and molecular gas at $\lesssim 10^{2.5}\,\mathrm{K}$.  
The term ``cloud--wind interaction'' and similar are used as a shorthand for
interactions between the clouds and hot wind components in the wind.
We stress, however, that the cloud particles interact with
all ambient gas that they encounter, not just wind material.
The label ``hydro'' is applied to quantities and measurements
related to the resolved gas treated by \textsc{Arepo}'s finite
volume scheme. This is often contrasted in figure legends
with ``Clouds'' which refers to gas modelled with cloud particles.
We adopt lowercase $r$ for a spherical radius, using $R$ for a
cylindrical radius in the plane of a galactic disc.

\section{Numerical Methods} \label{sec:methods}
\textsc{Arkenstone} is implemented in the \textsc{Arepo} code
(\citealt{Springel2010,Pakmor2016,Weinberger2020}).
In \cref{subsec:arepo} we describe relevant features of \textsc{Arepo} and other
details of the code setup used in this work that are not specific to
\textsc{Arkenstone}. In \cref{subsec:cloud_particle_ev}
we describe the new cloud particle scheme.
In \cref{subsec:launch_and_recouple} we describe the
launching and recoupling of wind and cloud particles,
briefly summarising relevant details of \textsc{Arkenstone-Hot}
that were presented in \citetalias{Smith2024}. 

\subsection{Hydrodynamics, gravity, cooling and the ISM} \label{subsec:arepo}
\textsc{Arepo} uses a finite volume scheme, solving 
hydrodynamics
on an unstructured, moving mesh. The mesh is defined by the Voronoi tessellation of
mesh-generating points which move with the local fluid velocity with small corrections
to maintain cell regularity. 
This means that cells tend to maintain constant mass over time,
giving the scheme quasi-Lagrangian properties.
However, while minimised by the mesh motion, mass fluxes
between cells are non-zero, so a (de)refinement scheme is typically used to (merge) split
cells to keep them within a factor of two of a desired mass resolution.
In addition to the constant mass (de)refinement scheme, other criteria can be used
to enforce varying mass or spatial resolution within the simulation domain.
\textsc{Arepo} can solve magnetohydrodynamics (MHD) \citep{Pakmor2011} 
but we do not include magnetic fields in this work.
None the less, we remark that \textsc{Arkenstone} is fully compatible with the
MHD scheme.
Gravity is included with a tree-based algorithm.\footnote{A TreePM scheme is available
but not used in this work.}

We include radiative cooling 
as described in \cite{Vogelsberger2013}. This includes cooling from both primordial
species \citep{Cen1992,Katz1996} and metal lines (in pre-calculated lookup
tables) in the presence of a $z=0$ UV background \citep{Faucher-Giguere2009},
with corrections for self-shielding in dense gas \citep{Rahmati2013}. 
While we do not impose a formal temperature floor, we do not radiatively cool
below $10^4\,\mathrm{K}$.
At the typical resolution at which \textsc{Arkenstone} is intended to operate, the multiphase
ISM cannot be well resolved. 
Therefore, in this work,
we use the model
of \cite{Springel2003}
(hereafter \citetalias{Springel2003}),
adopting an effective equation of state (eEoS)
to represent the large-scale impacts of small-scale
ISM physics (e.g. local stellar feedback, turbulence, molecular cloud
formation and destruction etc.) in an abstract manner.
The eEoS and star formation are switched on for gas denser than
a threshold value of $\rho_\mathrm{SF} / m_\mathrm{p} = 0.2\ \mathrm{cm^{-3}}$.
For the fine details of our parameter choices for this model (which
are essentially the same as the TNG suite),
see \citetalias{Smith2024}. Relevant for this work,
however, is that this ISM model predicts a star formation
rate (SFR), $\dot{m}_\star$, for each cell denser than $\rho_\mathrm{SF}$.
This rate is stochastically sampled to create star and wind
particles (as described in \citetalias{Smith2024}), as well
as cloud particles (as described in \cref{subsec:launch_and_recouple}).
We emphasise that \textsc{Arkenstone} is agnostic as to the choice
of subgrid ISM model or the method used to assign an SFR to each
cell, except that it assumes that
internal ISM structure is unresolved. In future,
we will explore alternative models to \citetalias{Springel2003}
\cite[such as that
proposed by][Hassan et al. in press, as calibrated from resolved star-forming, multiphase ISM simulations]{Ostriker2022,Kim2024}.

\subsection{Cloud particle evolution} \label{subsec:cloud_particle_ev}

We now describe the implementation of the cloud particle scheme.
We will first outline how \textsc{Arkenstone} implements cloud particle
and background gas interactions in general terms -- i.e. the parts of
the scheme that are independent of any particular theory or assumptions
of cloud--wind interactions. We then give specific details on how
we have implemented the \citetalias{Fielding2022}
cloud evolution model.

\subsubsection{General model}

%Throughout this work we denote a quantity associated with a cloud particle
%as $q_\mathrm{part}$, with an individual cloud as $q_\mathrm{cl}$ and with
%a gas cell as $q_\mathrm{cell}$. Fluxes are denoted as $\dot{q}$ and refer to
%a flux from a gas cell to a cloud particle unless otherwise specified.

A cloud particle is interpreted as carrying some number of identical clouds, $N_\mathrm{cl}$.
The constraint that the clouds must be identical (meaning that they have the same properties
as all other clouds carried by the particle in which they are hosted at all times) arises
because the trajectories of clouds with different properties will diverge. The cloud number
relates the masses of individual subgrid clouds, $m_\mathrm{cl}$,
to the mass of the cloud particle, $m_\mathrm{part}$:
\begin{equation}
N_\mathrm{cl} = \frac{m_\mathrm{part}}{m_\mathrm{cl}}. \label{eq:Ncl}
\end{equation}
$N_\mathrm{cl}$ is assigned to a cloud particle at its creation and remains constant throughout
its evolution.\footnote{While not currently implemented, any form of particle splitting or merging
would naturally also involve changing $N_\mathrm{cl}$ in order to conserve cloud number.
Likewise, while not examined in this work,
cloud evolution models could be easily implemented that involve splitting
or merging clouds within the particle, modifying $N_\mathrm{cl}$.
However, the constraint that all clouds within the particle are identical persists.
}
Note that we do not constrain $N_\mathrm{cl}$ to be an integer nor to be greater than unity.
Thus, an ensemble of cloud particles represents the population of individual clouds in
a statistical sense.

With the constraint that all clouds hosted by a given cloud particle are identical to each other,
we can use $N_\mathrm{cl}$ to relate all changes of conserved quantities of clouds to those of
the particle. For example, the rate of change of mass of the cloud particle is
\begin{equation}
\dot{m} = N_\mathrm{cl} \dot{m}_\mathrm{cl},
\end{equation}
where $\dot{m}_\mathrm{cl}$ is the rate of change of mass of an individual cloud in the particle.

The mass of a particle changes as a result of exchanges with the gas cell in which it is located.
These exchanges are bidirectional in the sense that the particle can be accreting
mass from the background gas at a growth rate $\dot{m}_\mathrm{grow}$ while it is simultaneously
losing mass to the background gas at a loss rate $\dot{m}_\mathrm{loss}$. 
The determination of $\dot{m}_\mathrm{grow}$ and $\dot{m}_\mathrm{loss}$ depends on the adopted
cloud--wind interaction model.
The net transfer rate
from cell to particle is therefore
\begin{equation}
\dot{m} = \dot{m}_\mathrm{grow} - \dot{m}_\mathrm{loss}.
\end{equation}
Therefore, at any given moment, a particle may be experiencing a net inflow or outflow of mass from the background
gas or no net mass transfer. However, it is important to note that $\dot{m} = 0$ does not necessarily mean there are
no net transfers of other quantities taking place. For example, the net rate of metal mass transfer from
cell to particle is
\begin{equation}
\dot{m}_Z = Z_\mathrm{cell}\dot{m}_\mathrm{grow} - Z_\mathrm{part}\dot{m}_\mathrm{loss}, \label{eq:mZ}
\end{equation}
where $Z_\mathrm{cell}$ is the metallicity of the cell and $Z_\mathrm{part}$ is the metallicity of the
particle (which is equal to the metallicity of the subgrid clouds). This means that even if the
net mass transfer is zero, if the particle and cell have different metallicities there will still be
a net transfer of metals (unless both $\dot{m}_\mathrm{grow}$ and $\dot{m}_\mathrm{loss}$ are zero).
In general, all transfers of passive scalars (e.g. individual metal species, tracer dyes etc.)
are handled using an equivalent version of \cref{eq:mZ}.

Momentum transfer rates from cell to particle are
\begin{equation}
\dot{\mathbf{p}} = \dot{\mathbf{p}}_\mathrm{trans} + \dot{\mathbf{p}}_\mathrm{drag}, \label{eq:pdot}
\end{equation}
where the first term captures momentum transfer associated with the mass transfers between the cell and the particle and the second term
allows for the inclusion of an additional
drag/ram pressure acceleration force (the specifics of which depend on the adopted cloud--wind interaction model).
By conservation of momentum, we can derive
\begin{equation}
\dot{\mathbf{p}}_\mathrm{trans} = \dot{m}_\mathrm{grow} \mathbf{v_\mathrm{cell}} - \dot{m}_\mathrm{loss} \mathbf{v_\mathrm{part}}, \label{eq:ptrans}
\end{equation}
where the cell and particle velocities
are $\mathbf{v_\mathrm{cell}}$ and $\mathbf{v_\mathrm{part}}$, respectively.
The resulting acceleration of the cloud particle due to this component of the momentum transfer is
\begin{equation}
\dot{\mathbf{v}}_\mathrm{trans} = \frac{\dot{\mathbf{p}}_\mathrm{trans} - \dot{m}\mathbf{v}_\mathrm{part}}{m_\mathrm{part}} = - \left(\mathbf{v}_\mathrm{part} -  \mathbf{v_\mathrm{cell}} \right) \frac{\dot{m}_\mathrm{grow}}{m_\mathrm{part}}. \label{eq:vdot}
\end{equation}
Note that this acceleration depends solely on $\dot{m}_\mathrm{grow}$, not on the net mass transfer.
Note also that the direction of this acceleration is always anti-parallel to the relative velocity between the particle and the cell,
$\mathbf{\mathbf{v}}_\mathrm{rel} = \mathbf{v}_\mathrm{part} -  \mathbf{v_\mathrm{cell}}$.
 Since this is also usually
true for definitions of the drag force, $\dot{\mathbf{p}}_\mathrm{drag}$, the momentum transfer between cell and particle always acts to reduce their relative
velocity, as might be intuitively expected.

We define the rate $\dot{E}_\mathrm{th}$
for the net transfer rate of thermal energy from the cell to the particle. Kinetic energy transfer is implicitly
mediated via the momentum transfer described above.

The source terms arising from the exchange of mass, momentum, energy and passive scalars (e.g. metals) are integrated in an explicit fashion by half timesteps immediately before and after the cloud particle receives its first and second gravity kicks, respectively.\footnote{Thus far,
we have found that timestep limiters required to integrate the \citetalias{Fielding2022} cloud--wind interaction model
(detailed below) with an explicit method are sufficiently computationally tractable that a semi-implicit or implicit scheme is not necessary.}
We limit the timestep of the cloud particle, $\Delta t$, such that:
\begin{equation}
\Delta t = \mathrm{MIN}\left(t_\mathrm{grav}, t_\mathrm{cell}, t_\mathrm{cellmod}, t_\mathrm{grow}, t_\mathrm{loss}, t_\mathrm{stop}, t_\mathrm{cross} \right),
\end{equation}
where each of these time-scales is defined below. 
The first timestep limit is the gravitational timestep limiter, $t_\mathrm{grav}$, which is the standard limiter for all particles (e.g. dark matter, stars, black holes)
and gas cells in the simulation. The default choice in \textsc{Arepo} is
\begin{equation}
t_\mathrm{grav} = \sqrt{\frac{2 C_\mathrm{grav} \epsilon_\mathrm{soft}}{\left | \mathbf{a} \right|}},
\end{equation}
where $C_\mathrm{grav}=0.012$ (the typical choice), $\epsilon_\mathrm{soft}$ is the gravitational softening length and $\left | \mathbf{a} \right|$
is the magnitude of the gravitational acceleration.

The timestep limit $t_\mathrm{cell}$ corresponds to the current timestep of the gas cell in which the cloud particle is located, which is itself
limited by the gravitational timestep limit, the hydrodynamical timestep limit (the Courant criterion) and any other limiters that may be
applied by additional physics.

We ensure that we resolve the timescale on which the host cell's properties are modified by enforcing the timestep limit
\begin{equation}
t_\mathrm{cellmod} = f_\mathrm{cellmod}\frac{m_\mathrm{cell}}{\left| \dot{m}\right|},
\end{equation}
where $f_\mathrm{cellmod}$ is a free parameter. For the case of net cloud mass growth,
this limiter ensures that the cell consumption time is resolved (in particular that
the particle will not attempt to accrete more mass than is available).
In the case of net cloud mass loss, this assists in allowing the cell's mass to grow smoothly,
rather than in a sudden injection of material. However, this latter case depends more
strongly on the timestep with which the cell is being integrated, which we discuss later in this section.

We also define timestep limits associated with the time-scales for mass growth and loss relative
to the cloud particle mass as follows:
\begin{equation}
t_\mathrm{grow} = f_\mathrm{grow}\frac{m_\mathrm{part}}{\dot{m}_\mathrm{grow}},
\end{equation}
\begin{equation}
t_\mathrm{loss} = f_\mathrm{loss}\frac{m_\mathrm{part}}{\dot{m}_\mathrm{loss}}, \label{eq:tloss}
\end{equation}
where $f_\mathrm{grow}$ and $f_\mathrm{loss}$ are free parameters. Note that we resolve the growth
and loss time-scales independently, rather than purely considering the net growth/loss
time-scale (which necessarily cannot be shorter than either the independent growth or loss time-scales).
This allows us to resolve changes in properties of the cloud particle that are correlated with
the fluxes in and out of the particle (e.g. metal transfer) even when the net mass transfer rate is
small (or zero).
It can be seen that $t_\mathrm{loss}$ can tend to zero when a cloud is being destroyed, which would
obviously be intractable to resolve. However, this is avoided by simply fully recoupling a cloud particle
once its mass drops below some threshold.

As noted above, all cloud - wind interactions reduce the relative velocity between the
cloud and the wind. We wish to resolve the stopping time of the cloud particle i.e. the
time for the relative velocity between the cloud particle and the cell to reach zero.
Failing to do so sufficiently results in spurious oscillations of the velocity of the cloud particle
as it repeatedly overshoots the velocity of the cell. 
We therefore define the timestep limit
\begin{equation}
t_\mathrm{stop} = f_\mathrm{stop} \frac{v_\mathrm{rel}}{\dot{v}_\mathrm{rel}} = f_\mathrm{stop} \frac{m_\mathrm{part} v_\mathrm{rel}}{\left| \dot{\mathbf{p}}\right| - v_\mathrm{rel} \dot{m}},
\end{equation}
where $f_\mathrm{stop}$ is a free parameter
and $v_\mathrm{rel}$ is the magnitude of the relative velocity.
Note that this estimate of the stopping time only accounts for changes in the relative velocity arising from \cref{eq:pdot}. However,
the gravitational acceleration will be approximately the same for the particle and the cell due to their proximity, so this will be
a minor contributor to changing $v_\mathrm{rel}$. The cell may feel additional forces that the particle does not (e.g. hydrodynamics)
but if these are dominant then this will already be captured via the $t_\mathrm{cell}$ timestep limiter.
It is possible for $t_\mathrm{stop}$ to tend to zero as $v_\mathrm{rel}$ tends to zero, depending on the dependence of $\dot{\mathbf{p}}$ on $v_\mathrm{rel}$. However, in our applications of the model so far, we have not encountered this problem.

Finally, we wish to resolve the crossing time of the cell by the particle to minimise the skipping of cells in its
path, which would otherwise happen for very high relative velocity between the background wind and the cloud particles.
We therefore define the timestep limit:
\begin{equation}
t_\mathrm{cross} = f_\mathrm{cross} \left(\frac{3 V_\mathrm{cell}}{4 \pi} \right)^{\frac{1}{3}} v_\mathrm{rel}^{-1},
\end{equation}
where $f_\mathrm{cross}$ is a free parameter and $V_\mathrm{cell}$ is the cell volume.

The timestep limits defined above are applied to the cloud particle. However, if it happens
that the host gas cell will be on a timestep longer than $t_\mathrm{cellmod}$, we move it
to a shorter timestep at the next available opportunity.\footnote{Within the current implementation of \textsc{Arepo}'s hierarchical timestep scheme, cells cannot be moved to a different timestep bin if
they are inactive. This is obviously a limitation as we cannot `wake up' a cell to guarantee that it is on an appropriate timestep. However, we have found empirically that it is rare for a cell to be on a timestep much longer than $t_\mathrm{cellmod}$ for an appropriate choice of $f_\mathrm{cellmod}$ as they are typically hot and spatially well refined, due to the \textsc{Arkenstone-Hot} scheme.}
This is primarily so that the radiative cooling of the cell can respond to changes in its properties. This approach works well and required no modifications to \textsc{Arepo}'s cooling routines, although in future we could instead choose to sub-cycle the cooling.
We conservatively adopt $f_\mathrm{cellmod} = f_\mathrm{grow} = f_\mathrm{loss} = f_\mathrm{stop} = 0.1$ and $f_\mathrm{cross}=0.3$ in this work,
finding this to provide accurate integration (we demonstrate this
in \cref{subsec:valid}) while still not contributing noticeably to
the cost of the simulation.
Except for the early stages of the cloud particle's evolution,
when the relative velocity with the ambient wind is at its highest,
cloud particles are typically limited by $t_\mathrm{grav}$.

\subsubsection{Implementation of the FB22 model}
\citetalias{Fielding2022} details a model for cloud--wind interactions that can be adopted by
our cloud particle scheme.
Clouds grow by accreting hot wind material via a turbulent radiative mixing layer (TRML) and lose
mass by turbulent shredding.
For the specifics of the model, its derivation and for physical interpretation
of its predictions, we refer the reader to \citetalias{Fielding2022} itself. Here, we give the essential
outline of the model and describe its implementation within our cloud particle scheme.

\citetalias{Fielding2022} assumes that the cloud is in pressure equilibrium with the ambient medium. We can therefore obtain the density contrast between a cloud and the ambient medium in terms of a specific internal energy contrast,
\begin{equation}
\chi \equiv \frac{\rho_\mathrm{cl}}{\rho_\mathrm{cell}} = \frac{u_\mathrm{cell}}{u_\mathrm{cl}}. \label{eq:contrast}
\end{equation}
Here, the density, $\rho_\mathrm{cell}$, and specific internal energy, $u_\mathrm{cell}$, of the host cell are known. The specific internal energy of the clouds within the cloud particle, $u_\mathrm{cl}$, is set by the assumption that the cloud is in thermal equilibrium with photo-heating from the UV background and/or local sources. While this could be determined on-the-fly, for simplicity, in this work we follow \citetalias{Fielding2022} and assume that the clouds have a temperature of $10^4\,\mathrm{K}$. Thus, the density of the clouds, $\rho_\mathrm{cl}$, can be determined from \cref{eq:contrast}. For clarity,
we emphasise that $\rho_\mathrm{cl}$ is the density of the material inside the identical subgrid clouds hosted within the cloud particle, not the total mass of clouds divided by the cell volume or any similar property.

The characteristic radius of any of the subgrid clouds hosted by the cloud particle is
\begin{equation}
r_\mathrm{cl} = \left(\frac{3 m_\mathrm{part}}{4 \pi \rho_\mathrm{cl} N_\mathrm{cl}} \right)^{\frac{1}{3}},
\end{equation}
where we have used $N_\mathrm{cl}$ to relate the particle mass to the mass of an individual cloud. The turbulent velocity within the TRML is taken to be
\begin{equation}
v_\mathrm{turb} = f_\mathrm{turb} v_\mathrm{rel}.
\end{equation}
Following \citetalias{Fielding2022} we adopt $f_\mathrm{turb} = 0.1$. We also need to determine the
cooling time within the TRML.
Using ``mix'' to denote gas properties
within the TRML, this is
\begin{equation}
t_\mathrm{cool}\equiv \frac{u_\mathrm{mix}}{n^2_{\mathrm{H,mix}} \Lambda'_\mathrm{mix}}.
\end{equation}
With hydrogen mass fraction $X_\mathrm{mix}$ the hydrogen number
density is
$n_\mathrm{H,mix}=X_\mathrm{mix}\rho_\mathrm{mix}/m_\mathrm{p}$.
$\Lambda'_\mathrm{mix} = \mathrm{max}\left[\Lambda(u_\mathrm{mix},\rho_\mathrm{mix},X_\mathrm{mix},Z_\mathrm{mix}),0\right]$ where $\Lambda$ is the net cooling rate
calculated using \textsc{Arepo}'s
cooling routines. $\Lambda'_\mathrm{mix}=0$ in the event of 
net heating (i.e. $\Lambda < 0.0$),
leading to an infinite cooling time.\footnote{In practice, to avoid potential floating point exceptions, if this occurs, we directly set $\xi$
(defined below) to zero.}
This does not ever occur in this work,
but might for lower choices
of $u_\mathrm{cl}$ and/or
in the presence of a strong
radiation field.
Following \cite{Begelman1990} and
\cite{Gronke2018}, and with the TRML in pressure equilibrium with
the cloud and the ambient medium
we have
 $u_\mathrm{mix} = \sqrt{u_\mathrm{cell}u_\mathrm{cl}}$,
 $\rho_\mathrm{mix} = \sqrt{\rho_\mathrm{cell}\rho_\mathrm{cl}}$,
$X_\mathrm{mix} = \sqrt{X_\mathrm{cell}X_\mathrm{cl}}$
and
$Z_\mathrm{mix} = \sqrt{Z_\mathrm{cell}Z_\mathrm{cl}}$.
As noted in \citetalias{Fielding2022},
what exactly sets $t_\mathrm{cool}$
in the TRML is still an open area
of research, so other choices
exist \citep[see e.g.][]{Abruzzo2022}.

We can then calculate the quantity
\begin{equation}
\xi = \frac{r_\mathrm{cl}}{v_\mathrm{turb}t_\mathrm{cool}},
\end{equation}
which compares the relative strengths of turbulent mixing and radiative cooling.
Referring the reader to \citetalias{Fielding2022} for the derivation, the mass growth
and loss rates are, respectively,
\begin{equation}
\dot{m}_\mathrm{grow} = \dot{m}_0 \xi^\alpha \label{eq:fb22_grow}
\end{equation}
and
\begin{equation}
\dot{m}_\mathrm{loss} = \dot{m}_0, \label{eq:fb22_loss}
\end{equation}
where
\begin{equation}
\dot{m}_0 = 3 f_\mathrm{mix} \frac{m_\mathrm{part} v_\mathrm{turb}}{\chi^{1/2}r_\mathrm{cl}},
\end{equation}
with $f_\mathrm{mix}=1/3$ our fiducial choice, and
\begin{equation}
\alpha = \begin{cases} 
1/4 & \xi \geq 1 \\
1/2 & \xi < 1
\end{cases}.
\end{equation}

These mass exchanges give rise to a momentum exchange, as expressed in \cref{eq:ptrans}. Additionally,
we define the drag term (which enters into \cref{eq:pdot}) as
\begin{equation}
\dot{\mathbf{p}}_\mathrm{drag} = -\frac{1}{2}C_\mathrm{drag}N_\mathrm{cl}\pi r_\mathrm{cl}^2\rho_\mathrm{cell}v_\mathrm{rel}\mathbf{v}_\mathrm{rel}, \label{eq:pdrag}
\end{equation}
where, as in \citetalias{Fielding2022}, we take $C_\mathrm{drag}=1/2$.

The mass exchanges also give rise to an exchange of enthalpy between the particle and the cell, such that the net flow of thermal energy from cell to particle is
\begin{equation}
\dot{E}_\mathrm{th} = \gamma \left(u_\mathrm{cell} \dot{m}_\mathrm{grow} - u_\mathrm{cl} \dot{m}_\mathrm{loss} \right). \label{eq:Edot}
\end{equation}
As previously mentioned, the temperature of the cloud particle is fixed; the specific thermal energy brought in by accreted material is radiated away in the TRML
(this condition has set $\dot{m}_\mathrm{grow}$ in the first place) so the cloud particle does not gain the energy described by \cref{eq:Edot}.
However, the gas cell loses this energy and also experiences an additional heating rate as the kinetic energy lost by the particle in the rest frame of the cell is thermalised. Thus, the thermal
energy of the cell changes by the following rate:
\begin{equation}
\dot{E}_\mathrm{cell,th} = \frac{1}{2} \dot{m}_\mathrm{loss} v_\mathrm{rel}^2 - \dot{E}_\mathrm{th}.
\end{equation}
Note that the total energy of the cell also changes as a result of the momentum transfers (including the drag force).

\subsubsection{Validation of the FB22 implementation} \label{subsec:valid}
\begin{figure*} 
\centering
\includegraphics{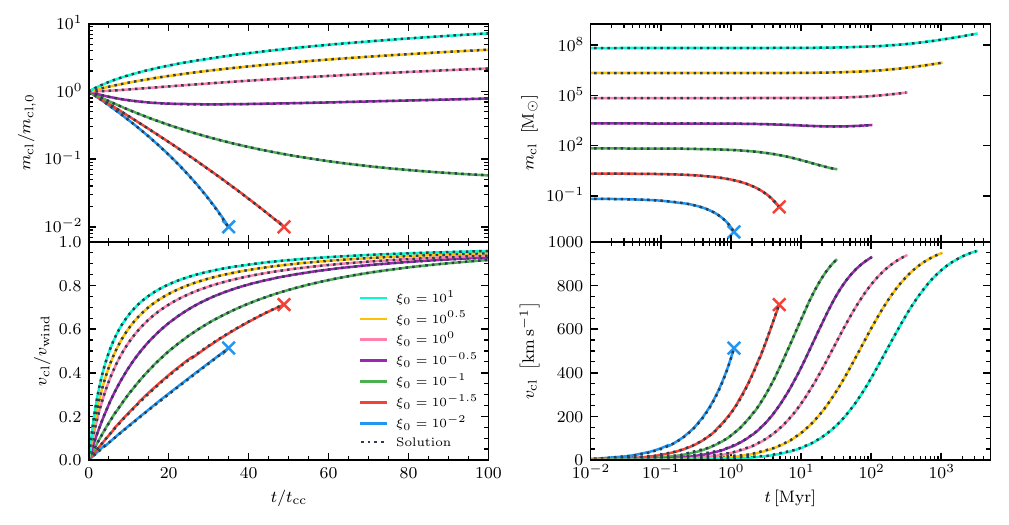}
\caption{The evolution of cloud particles placed in a wind tunnel, for various initial values of $\xi=\xi_0$ (corresponding to different initial cloud masses).
The wind has a velocity of $1000\ \mathrm{km\,s^{-1}}$, temperature of $10^7\ \mathrm{K}$ and pressure of $P/k_\mathrm{B} = 10^4\ \mathrm{K\,cm^{-3}}$. The density contrast between the clouds and the wind is 1000. To enable a one-to-one
comparison to the \citetalias{Fielding2022} solutions we adopt a fixed value of $t_\mathrm{cool}=1\ \mathrm{Myr}$.
Solid lines indicate the results from the \textsc{Arkenstone} simulations while the dotted lines indicate the 
directly integrated solutions from \citetalias{Fielding2022}.
We run the simulations for 100 initial cloud crushing times ($t_\mathrm{cc}$) or until the particles lose 99 per cent of their mass. The latter case is indicated by a cross marker.
\textit{Top left}: cloud mass normalised to the initial cloud mass as a function of time normalised to $t_\mathrm{cc}$.
\textit{Bottom left}: cloud velocity normalised by the wind velocity as a function of time normalised to $t_\mathrm{cc}$.
\textit{Top right}: cloud mass as a function of time.
\textit{Bottom right}: cloud velocity as a function of time.
It can be seen that the implementation of the \citetalias{Fielding2022} cloud evolution model into \textsc{Arkenstone} results in essentially perfect agreement with the directly computed solutions across a wide range of parameters.
}
\label{fig:windtunnel} 
\end{figure*}

In order to validate our implementation, we present a simple numerical experiment. A single cloud particle is placed in a hot wind tunnel
(similar in principle to many cloud--crushing simulations) and its subsequent evolution is
compared to the \citetalias{Fielding2022} solution. We choose the same configuration as \citetalias{Fielding2022},
section 3.2.3 (the results of which are shown in their fig. 4). The background wind has a velocity of
$1000\ \mathrm{km\,s^{-1}}$ and a temperature of $10^7\ \mathrm{K}$.
Both the wind and clouds have a pressure of $P/k_\mathrm{B} = 10^4\ \mathrm{K\,cm^{-3}}$ and we use $\chi=1000$.
To avoid having to implement \textsc{Arepo}'s cooling functions into our separate routines for directly
integrating the \citetalias{Fielding2022} model (or vice-versa), for this specific test we choose to adopt a fixed value of
$t_\mathrm{cool} = 1\ \mathrm{Myr}$ (which is of the correct order of magnitude for the cooling functions in the regime probed)
to enable an exact evaluation of the performance of \textsc{Arkenstone} with respect to the \citetalias{Fielding2022} solutions.
The background gas is not permitted to radiatively cool (other than onto the cloud particle) as a constant background
is necessary for comparison to the idealised solutions. Likewise, we do not include any self-gravity.
We perform several simulations with different initial values of $\xi$ for the cloud particles, each of which specifies
an initial cloud mass (or, equivalently, an initial cloud radius).

We choose a target gas cell resolution of $8\times10^4\ \mathrm{M_\odot}$ for the background wind. The domain has dimensions of
$100\ \mathrm{kpc}\,\times\,25\ \mathrm{kpc}\,\times\,25\ \mathrm{kpc}$ with periodic boundary conditions. Initial positions of the
mesh generating points are drawn from a low discrepancy sequence \citep[the $R_3$ sequence of][]{Roberts2018} in order to
reduce Poisson noise while avoiding a highly structured configuration as much as possible. Cell masses are initialised by multiplying
the target density with their initial volume so that any remaining noise in the initial mesh generating point configuration manifests
as a scatter in cell mass rather than density. This configuration is then evolved for several domain sound crossing times
with \textsc{Arepo}'s standard refinement, de-refinement and mesh regularisation schemes enabled, yielding a completely uniform density
medium represented with an unstructured mesh and well-rounded cells which all have masses within the standard \textsc{Arepo} tolerance of a factor
of 2 of the target resolution. The cloud particle is given an initial mass of $800\ \mathrm{M_\odot}$
(not to be confused with the initial subgrid cloud masses), placed in the
domain and given an initial velocity of $1000\ \mathrm{km\,s^{-1}}$ (thus, in the rest frame of the particle, it is experiencing
a $1000\ \mathrm{km\,s^{-1}}$ wind). We run the simulations for 100 cloud crushing times, where $t_\mathrm{cc} = \chi^{1/2} r_\mathrm{cl} / v_\mathrm{rel}$, or
until the cloud particle has lost 99 per cent of its initial mass.

\cref{fig:windtunnel} shows the results of these simulations. We show the evolution of the cloud mass and velocity both in absolute terms or normalised to the initial mass or
wind velocity, respectively. The solid, coloured lines show the output of the simulations while the dotted lines show the
direct integral of the \citetalias{Fielding2022} equations using a standard RK45 integration scheme.\footnote{The
solutions that we compare against neglect the backreaction of the cloud
on the wind while this is included in our simulations. However, the impact occurs
predominantly downstream and is never felt by the single cloud particle used in this test, which
moves onwards into a reservoir of pristine gas. If we switch off the backreaction
terms for this test, we produce essentially identical results.}
It can be seen
that for a range of initial values of $\xi=\xi_0$, the simulations have essentially perfect agreement with the expected evolution, confirming that
our implementation correctly integrates the cloud evolution mass, momentum and energy transfer rates.
For a detailed physical interpretation
of the behaviour of the cloud evolution model, we refer the reader to \citetalias{Fielding2022}. However, the experiment
highlights the most salient points.
Larger values of $\xi_0$ correspond to initially larger
and therefore (for fixed $\chi$) more massive clouds.
Large clouds (with $\xi_0\gg1$) grow efficiently and
are accelerated to a significant fraction of the wind
velocity within a few $t_\mathrm{cc}$.
Small clouds (with $\xi_0\ll1$) lose mass quickly.
Small clouds evolve much faster than the larger clouds
in real terms, undergoing a more rapid acceleration
(despite the acceleration time-scale being longer
relative to their own $t_\mathrm{cc}$ compared to
larger clouds).
Intermediate clouds with $\xi_0 \lesssim 1$
lose mass to begin with, but the instantaneous
$\xi$ becomes greater than unity as the cloud
is entrained in the wind ($v_\mathrm{rel}$ decreases)
leading to subsequent net growth.

\subsection{Launching and recoupling of cloud and wind particles} \label{subsec:launch_and_recouple}

In this work, we only consider the generation of clouds from within the ISM
(rather than via precipitation from the CGM for example, although the scheme can also be applied in that context).
In \citetalias{Smith2024} we described the launching of wind particles from the ISM in order to drive the
hot phase of a galactic wind - the \textsc{Arkenstone-Hot} scheme. Cloud particles are launched following the same approach. We therefore refer the interested reader to \citetalias{Smith2024}, but summarise the details here, focusing primarily on the addition of the cloud particles to the scheme.

The star formation rate of a cell, $\dot{m}_\star$, is calculated in this work with the model of \citetalias{Springel2003}, though our scheme is insensitive to this choice. The input wind and cloud mass loading factors, $\eta_{M}^{\mathrm{w}0}$ and $\eta_{M}^{\mathrm{cl}0}$,
relate the rate at which wind and cloud material is injected relative to the star formation rate:
\begin{equation}
\dot{m}_\mathrm{w} = \eta_{M}^{\mathrm{w}0} \dot{m}_\star,
\end{equation}
\begin{equation}
\dot{m}_\mathrm{cl} = \eta_{M}^{\mathrm{cl}0} \dot{m}_\star,
\end{equation}
We sample $\dot{m}_\star$, $\dot{m}_\mathrm{w}$ and $\dot{m}_\mathrm{cl}$ to stochastically generate star, wind and cloud particles from star forming gas cells.
Star particles have the same mass resolution as the gas mass resolution of the simulation (so gas cells are typically converted completely into star particles), whereas we spawn wind and
cloud particles that have a mass that is a factor $f_{m}^{\mathrm{w}}$ and $f_{m}^{\mathrm{cl}}$, respectively,
smaller than the gas mass resolution.

Wind particles inherit the velocity of their parent gas cell, but are given a kick
\begin{equation} \label{eq:vwind}
\Delta v_\mathrm{w} = \sqrt{\frac{2\eta_{E}^{\mathrm{wkin}0}}{\eta_{M}^{\mathrm{w}0}} u_\star},
\end{equation}
where $\eta_{E}^{\mathrm{wkin}0}$ is the input wind kinetic energy loading and $u_\star$ is the characteristic specific energy associated with stellar feedback. 
Since the majority of the energy driving the winds originates from supernovae (while radiation, stellar winds etc. have a more local effect around massive stars),
we adopt $u_\mathrm{\star} = 5.26\times10^5\ (\kms)^2$, as in
\citetalias{Smith2024}, corresponding to one SN of
10$^{51}$~erg for every 95.5~$\Msun$ of stellar mass formed (consistent
with the value used in \citealt{Kim2020b}).\footnote{It should be noted that $u_\star$ plays the role of a normalising reference value and is degenerate with the energy loadings. Its exact value
is not particularly important, except when comparing energy loadings between works.}
The kick can be applied in some preferred direction (e.g. vertically out of the disc
plane) or isotropically.
This choice is discussed in detail in \citetalias{Smith2024}.
For this work we choose to apply the kicks vertically.
The wind particle receives a specific internal energy
\begin{equation} \label{eq:uwind}
u_\mathrm{w} = \frac{\eta_{E}^{\mathrm{wth}0}}{\eta_{M}^{\mathrm{w}0}} u_\star,
\end{equation}
where $\eta_{E}^{\mathrm{wth}0}$ is the input wind thermal energy loading.

In this work we choose to set
the initial velocity and internal energy of the cloud particles directly (i.e. this means they are independent of the mass loading). 
The specific internal energy of the clouds are set by the cloud model.
As described above, for the \citetalias{Fielding2022} model adopted in this work,
this means that we give the clouds an initial temperature of $10^4\,\mathrm{K}$. 
Note that we cannot resolve
the acceleration within the ISM, so the choice of the initial launch velocity is intended to compensate for this.
In the future, we will derive this value more rigorously from high resolution simulations in tandem with
an appropriate mass loading for cool gas that is able to escape the ISM (possibly leading to velocities that depend on halo properties).
However, in this work, we choose a value of $\Delta v_\mathrm{cl} = 100~\kms$ which we have determined empirically
to be high enough to allow all cloud particles to escape the ISM in our idealised setup while still
providing plenty of headroom for the interaction with the hot wind to accelerate them to higher velocities.
We will demonstrate later in this work that this initial velocity kick is negligible compared to the subsequent cloud
acceleration by the hot wind.

Both wind and cloud particles are initially hydro-decoupled, meaning that they only experience gravity and do not participate in any hydrodynamical interactions (either explicitly or via a subgrid model). Once
a wind particle finds itself in gas with a density lower than a threshold, $\rho_\mathrm{rec}$, it recouples, depositing its mass, energy, momentum and metals
into the gas of the ISM/CGM transition region and drives a hydro-resolved hot wind. 
We choose $\rho_\mathrm{rec} = 0.1\rho_\mathrm{SF}$.
The details of the \textsc{Arkenstone-Hot} wind recoupling model
are described in detail in \citetalias{Smith2024}. The most salient point, where we have made significant improvements
over previous similar models, is that we are able to ensure that we have sufficient resolution to resolve a low density, high
specific energy outflow with the use of a novel refinement technique (``displacement recoupling''). The wind is initially kept
at the same mass resolution
as the wind particles (i.e. a factor $f_{m,\mathrm{w}}$ better than the base resolution of the
simulation). Wind material subject to the refinement criteria is identified with a passive scalar "dye"
injected as the wind particle recouples. In this work, we gradually relax the refinement of the wind such that
the mass resolution increases linearly from $0.1r_{200}$ from the galaxy centre until it reaches the base resolution of the simulation
at $0.5r_{200}$. A more physically motivated version of this distance--refinement level relation implemented for
use in cosmological simulations will be presented in Bennett et al. in prep.

If the ambient density around a cloud particle is greater than $\rho_\mathrm{rec}$ and the density contrast
between the clouds and the ambient medium is smaller than a threshold value $\chi_\mathrm{dec}$, it also remains
completely hydro-decoupled. The first time
that a cloud particle finds itself in gas such that $\rho_\mathrm{cell} \leq \rho_\mathrm{dec}$ and
$\chi \geq \chi_\mathrm{dec}$, the full cloud evolution model is enabled and exchanges of mass, momentum, energy
and metals between the cloud and the host cell begin, as described in the previous sections. Once the
cloud model has been enabled, if the density contrast ever falls below a second threshold value, $\chi_\mathrm{rec}$,
or its mass drops below a factor $f_{m,\mathrm{rec}}$ of its initial value, it is fully recoupled into the cell.
This proceeds via the ``standard recoupling'' scheme described in \citetalias{Smith2024}; all conserved quantities
of the cloud particle are injected into the host cell. 
We adopt $f_{m,\mathrm{rec}} = 0.1$.
We choose $\chi_\mathrm{rec}=10$, causing the cloud particle
to recouple if it has encountered gas with densities not too dissimilar to the subgrid internal cloud density.
We use a slightly higher value to enable
the cloud evolution model,
choosing $\chi_\mathrm{dec}=1.1\chi_\mathrm{rec} = 11$,
such that cloud particles are sufficiently
clear of the recoupling threshold to avoid ``false starts''.

As described below, in this work all cloud particles are launched with sufficient
velocity to exit the ISM and meet the conditions to turn on the cloud evolution model.
In other settings (such as cosmological simulations), it is in principle possible for particles
to remain trapped in the ISM in which case we will fall back to a maximum lifetime before
recoupling. However, because our model is designed to treat winds that have left the ISM,
the launching of cloud particles with low velocities that will always remain
fully hydro-decoupled should be avoided as much as possible.
When measuring outflow properties
from high resolution simulations for use as inputs to \textsc{Arkenstone},
the mass loading of the cool phase should only
consider contributions from material that has a high enough velocity to leave the ISM.

\section{Demonstration of the model} \label{sec:results}{}
We will now apply the model using idealised non-cosmological simulations. The aim of this work is to demonstrate the general behaviour of the model in various regimes, rather than to advocate for a particular set of loading factors. A more comprehensive study of the effect of different loading factors will be the focus of future works in a cosmological context.

\subsection{Initial conditions and model setup}
\begin{table}
\caption{The parameters defining the initial conditions
used in this work. For details about the models adopted and the
definition of the symbols, see the
main text. Note that the quantities reported in this table are
all input parameters with the exception of $r_{200}$,
which is derived from the halo mass, concentration and cosmology,
and the CGM mass inside $r_{200}$, which is a derived
quantity of the cooling flow solution given the other constraints.}
\label{tab:IC}
\begin{center}
\begin{tabular}{lr}
\hline
Parameter & Value\\
\hline
\textbf{Dark matter} &\\
$M_{200}$ & $10^{11}\,\Msun$\\
$c$         & 10\\
$r_{200}$ & 97.9 kpc\\
$s_e$       & 1.5\\
$b_e$       & 1\\
\hline
\textbf{Stellar disc} &\\
$M_{\mathrm{disc},\star}$   & $8\times10^{9}\,\Msun$\\
$R_\mathrm{s}$ & 2.5 kpc\\
$z_\mathrm{s}$ & 0.25 kpc\\
$m_\star$      & $8\times10^{4}\,\Msun$\\
\hline
\textbf{Stellar bulge} &\\
$M_{\mathrm{bulge},\star}$    & $10^{8}\,\Msun$\\
$r_\mathrm{s}$ & 2.5 kpc\\
$m_\star$      & $8\times10^{4}\,\Msun$\\
\hline
\textbf{Gas disc} &\\
$M_{\mathrm{disc},\mathrm{gas}}$    & $2\times10^{9}\,\Msun$\\
$R_\mathrm{s}$ & 2.5 kpc\\
$T_\mathrm{0}$ & $10^4\,\mathrm{K}$\\
$Z_\mathrm{0}$ & $1\,Z_\odot$\\
$m_\mathrm{g,tar}$ & $8\times10^{4}\,\Msun$\\
\hline
\textbf{CGM}&\\
$r_\mathrm{circ}$ & 2.5 kpc\\
$r_\mathrm{sonic}$ & 2 kpc\\
$Z_{0}$ & $0.1\,Z_\odot$\\
$M_\mathrm{CGM}\left(<r_{200}\right)$ & $2.96\times10^9\,\Msun$\\
$m_\mathrm{g,tar}$ & $8\times10^{4}\,\Msun$\\
\hline
\end{tabular}
\end{center}
\end{table}

We use the fiducial initial conditions from \citetalias{Smith2024} and refer the interested reader there for the finer details of their preparation. The setup is an isolated system comprised of dark matter, a disc and bulge of pre-existing stars, a gas disc and a realistic CGM.
The latter component is often omitted
from idealised galaxy simulations but is
crucial for this work, allowing
us to study accretion onto the galaxy,
the interaction of winds with
the CGM and the balance of
ejective and preventative feedback.
The parameters describing the initial conditions are given for reference in \cref{tab:IC}.
The dark matter is modelled as a spherically symmetric, static background potential and includes both an
inner and outer halo component. The inner component follows a Navarro-Frenk-White \citep[NFW,][]{Navarro1997}
profile with $M_{200}=10^{11}\,\Msun$
and a concentration of 10. Taking a
\cite{PlanckCollaboration2020} cosmology, this gives $r_{200}=97.9\,\mathrm{kpc}$ (where we define
$M_{200}$ and $r_{200}$ relative to the critical
density at $z=0$).
The outer halo component is modelled following \cite{Diemer2014}, with the parameters $s_e=1.5$ and $b_e=1$. The stellar disc and bulge, and the gas disc are generated using
the code \textsc{MakeNewDisk} \citep{Springel2005}. The stellar and gas discs have exponential
surface density profiles with a scale length of $R_\mathrm{s}=2.5\,\mathrm{kpc}$. The stellar disc has a Gaussian vertical
density profile with a scale height of $z_\mathrm{s}=0.25\,\mathrm{kpc}$. The gas disc has a vertical density profile that
is set to produce hydrostatic equilibrium at its initial temperature of $T_0=10^4\,\mathrm{K}$. 
We truncate the gas disc beyond five scale lengths and five scale heights.
The disc has an initial metallicity of $Z_{0}=1\,Z_\odot$ (where we adopt $Z_\odot=0.0127$).
The stellar bulge is spherically symmetric and follows a \cite{Hernquist1990} density profile
with a scale length of $r_\mathrm{s}=0.25\,\mathrm{kpc}$. We use
$M_{\mathrm{disc},\star}=8\times 10^9\,\Msun$, $M_{\mathrm{bulge},\star}=10^8\,\Msun$ and $M_\mathrm{disc,gas}=2\times 10^9\,\Msun$.

The CGM gas is initialised to a steady state rotating cooling flow configuration for our total potential, the full details of which can be found in \cite{Stern2024}. We choose a cooling flow solution with a sonic radius of the flow of $r_\mathrm{sonic} = 2\ \mathrm{kpc}$ and a circularisation radius of $r_\mathrm{circ} = 2.5\ \mathrm{kpc}$.
This means that the flow remains in the subsonic limit. We choose an initial CGM metallicity of $0.1\,Z_\odot$.
In combination with the other constraints, this yields an initial CGM mass inside $r_{200}$ of $2.96\times10^9\ \mathrm{M_\odot}$.
The predicted steady-state mass flux of the cooling flow (in the absence of feedback) for these initial conditions is $0.2\,\Msun\,\mathrm{yr}^{-1}$, though in practice
the addition of a galaxy to the centre of the halo and minor discrepancies between the cooling functions
used in \textsc{Arepo} and those assumed when calculating the solution leads to the emergent
inflow rate being a factor of a few higher.
In order to provide a reservoir of mass that can flow into the halo and maintain the cooling
flow, we initialise the CGM gas out to a distance of $600\ \mathrm{kpc}$ ($\sim 6r_{200}$). However, we do
not require our full resolution far outside the halo.
We therefore degrade the mass resolution smoothly outside $200\ \mathrm{kpc}$ by a factor of 3 every $\sqrt{2}\times200\ \mathrm{kpc}$, making appropriate modifications to the (de)refinement scheme to maintain this resolution
structure during the simulation.

We use a target gas cell mass resolution of $m_\mathrm{g,tar}=8\times10^4\ \mathrm{M_\odot}$. Recall that, as described above, a given gas cell may be at a finer resolution if subject to the \textsc{Arkenstone} hot wind refinement scheme or
a coarser resolution if beyond $200\ \mathrm{kpc}$ from the system centre. Star particles, either present in the
initial conditions or created during the simulation, also have a mass of $m_\star=8\times10^4\ \mathrm{M_\odot}$. Gas cells
have an adaptive gravitational softening of 2.5 times the cell radius with a minimum value of $50\ \mathrm{pc}$.
Collisionless particles have a fixed softening of $195\ \mathrm{pc}$.

For the background hot wind, we take the input loading parameters for the high specific energy wind presented in \citetalias{Smith2024}, $\eta_{M}^{\mathrm{w}0} = 0.32$, $\eta_{E}^{\mathrm{wkin}0}=0.321$, $\eta_{E}^{\mathrm{wth}0}=0.579$. We use $f_{m}^{\mathrm{w}}=0.01$ i.e. the wind particles and refined wind gas are 100 times less massive than the base resolution of the simulation. For the cool cloud component, we perform simulations with three different input mass loadings, $\eta_{M}^{\mathrm{cl}0} = 0.1$, 1 and 5, which illustrate different regimes of the model. As stated earlier, in this work we give cloud particles an initial velocity kick of $100~\kms$ and temperature of $10^4\,\mathrm{K}$. For the three mass loadings above, this corresponds to total input cloud energy loadings of $\eta_{E}^{\mathrm{cl}0}=9.7\times10^{-4}$, $9.7\times10^{-3}$ and 0.048, respectively. 
Note that these are negligible compared to the input energy loadings of the
hot wind.
We use $f_{m}^{\mathrm{cl}}=0.01$ i.e. cloud particles are initially 100 times lower mass than the base resolution of the simulation (but the same mass resolution as the wind particles and refined wind gas).
Newly created cloud particles are assigned a randomly drawn cloud number, $N_\mathrm{cl}$, such that the initial distribution of (subgrid) cloud masses across the whole population of cloud particles follows a mass function $\mathrm{d}N/\mathrm{d}m \propto m^{-2}$, motivated by the mass functions
seen in the high resolution simulations of \cite{Tan2024}.
We also perform simulations with no cool cloud component (i.e. $\eta_{M}^{\mathrm{cl}0} = 0$) and no wind at all (i.e. $\eta_{M}^{\mathrm{w}0}=\eta_{M}^{\mathrm{cl}0} = 0$).

For the sake of simplicity, in all simulations in this work, wind and cloud particles inherit the metallicity of the ISM gas cell from which they were spawned. We will include differential levels of enrichment in future work
\citep[derived from high resolution simulations such as][]{Kim2020b},
which we expect to have significant impact on the metallicity of different CGM phases. 

\subsection{Results}
\subsubsection{Morphologies}
\begin{figure*} 
\centering
\includegraphics{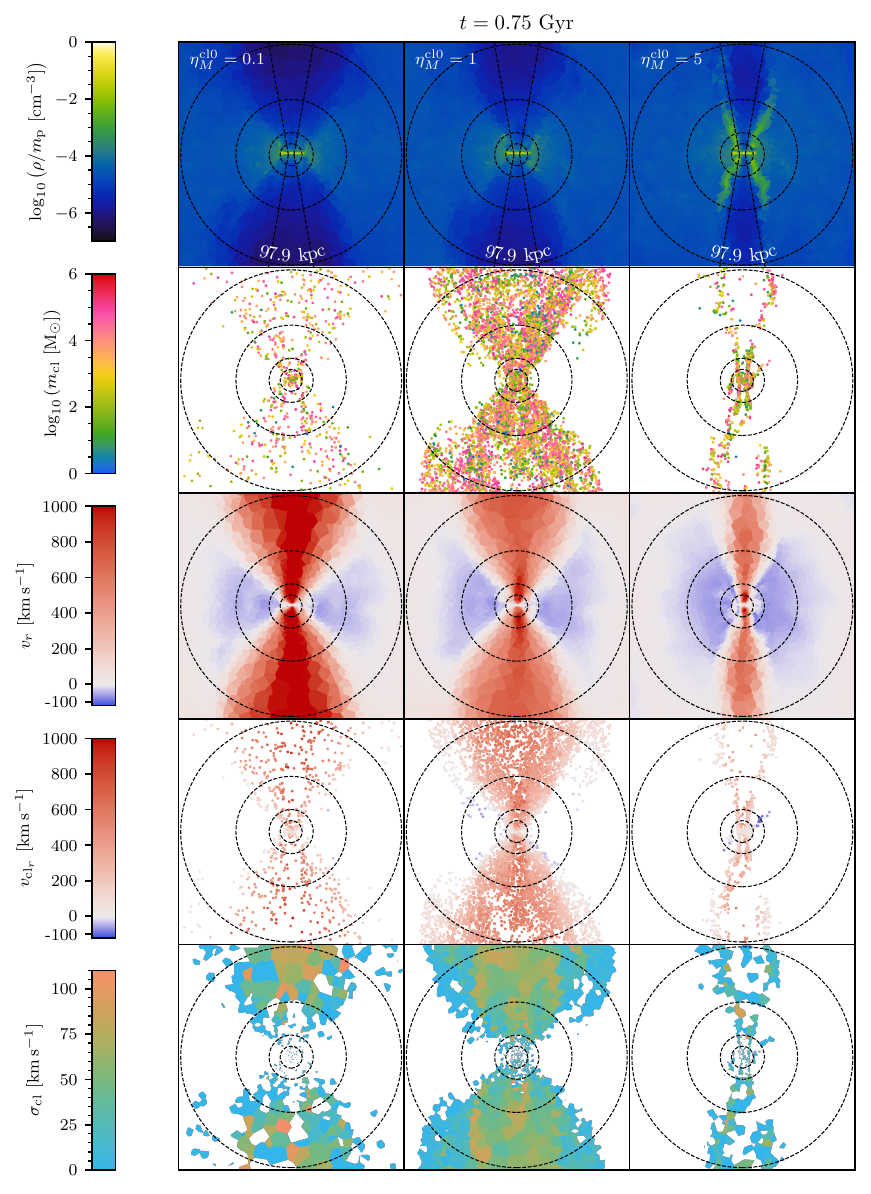}
\caption{Visualisations of the fiducial simulations with cloud particles after 0.75 Gyr.
The left, centre and right columns show the $\eta_{M}^{\mathrm{cl}0} = 0.1$, 1 and 5 simulations, respectively.
The top row shows slices of gas density, orientated vertically through the disc plane.
The second row shows the locations of all cloud particles contained within cells intersected
by the slice shown in the top row, coloured by their current subgrid cloud mass.
The third and fourth rows show the radial velocity for gas and cloud particles, respectively.
The fifth row shows slices of the intra-cell cloud particle 3D velocity dispersion (see the text
for details).
On all panels, we overlay dashed circles corresponding to 0.1, 0.2, 0.5 and 1 times $r_{200}$.
These are the reference surfaces used later in the work to measure mass and energy fluxes.
On the top row, we also indicate (again with dashed lines) the region used when measuring
various wind property profiles later in the work. The three simulations shown demonstrate different scenarios. On the left, a low cloud mass loading means that the background wind experiences a minimum of disruption, but there is only a small population of cloud particles. In the centre, a higher mass loading results in only a slight impact on the wind, but the wind is now filled with cloud particles. On the right, a mass loading of 5 has lead to significant disruption of the background wind, which has in turn limited the spatial extent covered by cloud particles. In the bottom row, we can see that the different trajectories of cloud particles means that the cloud velocity field would be poorly described by a single value per cell (which would enforce $\sigma_\mathrm{cl} = 0$).
}
\label{fig:slice_mvar_153} 
\end{figure*}

\begin{figure*} 
\centering
\includegraphics{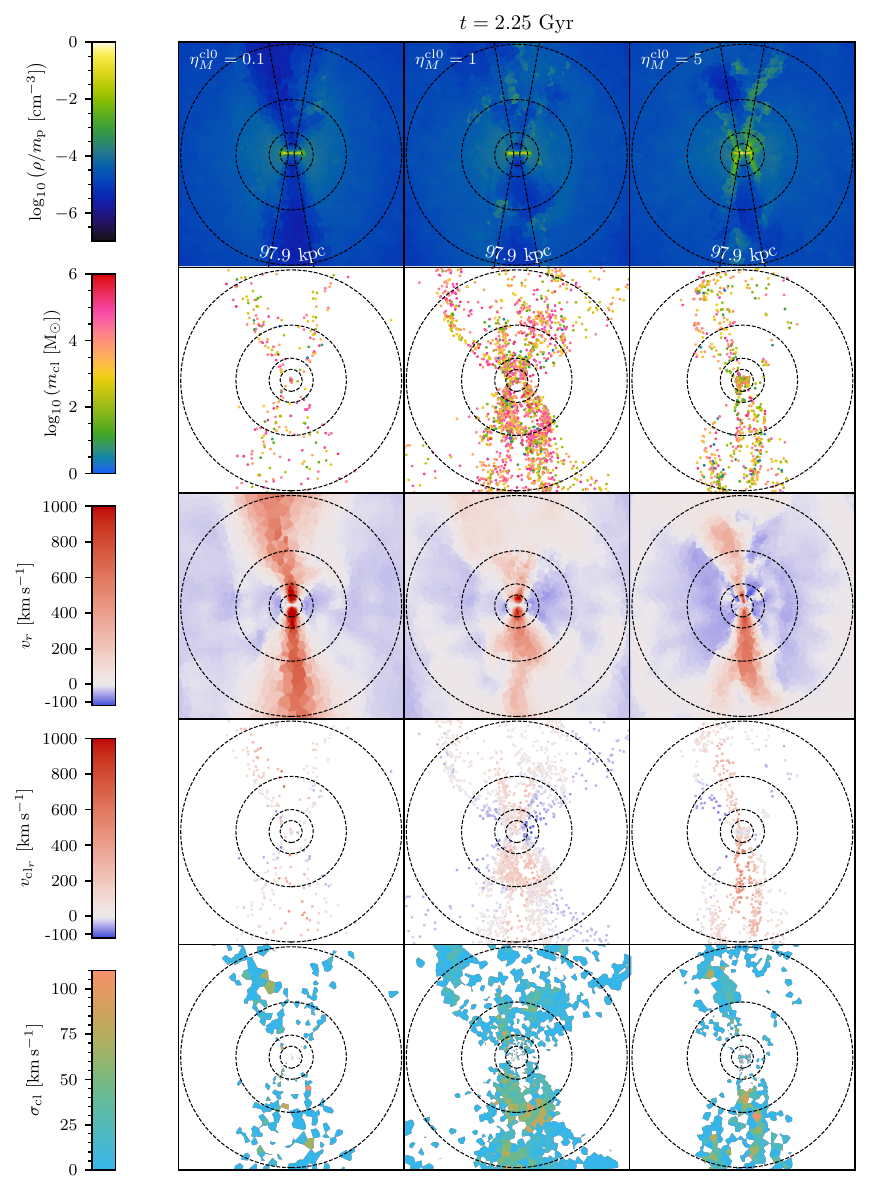}
\caption{Visualisations of the fiducial simulations with cloud particles after 2.25 Gyr. See the caption
for \cref{fig:slice_mvar_153} for details on what is displayed.
The bicone of the outflow in the $\eta_{M}^{\mathrm{cl0}}=0.1$ simulation (left) has narrowed but is largely undisrupted. Both the $\eta_{M}^{\mathrm{cl0}}=1$ and 5 (centre and right, respectively) show significant disruption
to the wind bicone, with the outflow being diverted by denser gas structures. This is more severe in the $\eta_{M}^{\mathrm{cl0}}=5$ case. The complex velocity field of the background wind (which also varies
significantly in time) leads to cloud particles falling out of the flow. There are therefore populations of inflowing cloud particles.}
\label{fig:slice_mvar_460} 
\end{figure*}

In \cref{fig:slice_mvar_153,fig:slice_mvar_460} we show visualisations of the simulations with cloud particles after 0.75 and 2.25 Gyr, respectively. In the top row, we show a gas density slice\footnote{The ``slice''  is a map of the properties of cells intersected by the image plane.}
orientated
vertically through the disk plane. 
The underlying structure of the Voronoi mesh is apparent in these slices; recall that these simulations are carried out at a resolution that is feasible for cosmological volume simulations, not just for individual galaxies.
Below this, we plot the locations of cloud particles coloured by
their current subgrid cloud mass. The particles shown are all those that are located inside the cells
which are shown in the gas density slice. The third and fourth rows show the radial velocity of the gas
and cloud particles, respectively. The bottom row shows a slice of the intra-cell cloud particle
3D velocity dispersion, $\sigma_\mathrm{cl}$. For each cell that contains cloud particles, we calculate
$\sigma_\mathrm{cl}$ as the mass-weighted velocity dispersion of all of their cloud particles.

Broadly speaking, \cref{fig:slice_mvar_153,fig:slice_mvar_460} neatly illustrate three distinct scenarios for the impact of the cloud population on the background wind as the input cloud mass loading is varied: 1) there are relatively few clouds in the outflow so the background wind is barely impacted by the cloud component, 2) there are large amounts of cloud material in the outflow but the background wind is still coherent and energetic, and 3) a high cloud mass loading significantly disrupts the background wind which in turn limits the spread of cloud material within the CGM. We will now explore these results in more detail.

Starting with the $\eta_{M}^{\mathrm{cl0}}=0.1$ simulation (left-hand columns), in \cref{fig:slice_mvar_153} (at 0.75 Gyr), the outflowing wind can be clearly seen in the density and radial velocity slices. The outflow forms a
coherent bicone that is significantly under-dense compared to the CGM. The outflow is fastest
up the centre of the bicone. The outflow covers a significant portion of the CGM, restricting inflows (and generally regulating the galaxy evolution as described in \citetalias{Smith2024}). By 2.25 Gyr (see \cref{fig:slice_mvar_460}) the opening angle of the outflow has reduced but is still a coherent flow. This effect was seen and discussed in detail in \citetalias{Smith2024} in the absence of cloud material. As the SFR of the galaxy drops (as we will examine below), the
power of the wind similarly drops. This allows the inflowing CGM to confine the outflow, leading to
a narrowing of the wind bicone. In both \cref{fig:slice_mvar_153,fig:slice_mvar_460}, a handful of
cloud particles can be seen with a wide variety of cloud masses. At 0.75 Gyr, these particles largely trace the wind region and are travelling outwards at many hundreds of $\mathrm{km\,s^{-1}}$ (we will examine the outflow velocities quantitatively later in the work), much faster than their initial launch velocity. 
Cloud material is able to make it to $r_{200}$ and beyond. It can be seen that there is a slight under-density in cloud particles in the centre of the outflow, where the velocity of the wind is highest. By 2.25 Gyr, there are fewer cloud particles and they are more preferentially located closer to the boundary of the wind. In fact, some are actually located outside of the wind. This arises because cloud particles are not perfectly
``pinned'' to the wind material. The wind velocity field can fluctuate faster than the clouds can
respond, due to their inertia, so clouds can fall out of the flow. Furthermore, the wind is in general faster than the clouds so as the wind bicone narrows, cloud particles can find themselves outside of the wind. If the clouds survive the sudden change in relative velocity as they pass into oncoming CGM material, they can then fall back towards the galaxy, as can be seen for some particles.

At 0.75 Gyr (\cref{fig:slice_mvar_153}), it can be seen that the hot wind in the $\eta_{M}^{\mathrm{cl0}}=1$ simulation (centre columns) is similar to the 
$\eta_{M}^{\mathrm{cl0}}=0.1$ case, albeit slower. It still occupies a large fraction of the CGM.
Many more cloud particles are present, spanning the wind bicone. Again, there is a wide variety of cloud masses (we will examine the mass distribution in a quantitative manner below) and cloud material
is able to be accelerated out of the halo. Already at this time, small amounts of inflowing cloud material is apparent. These are cloud particles launched at earlier times that have ended up outside of the wind. By 2.25 Gyr (\cref{fig:slice_mvar_460}), the wind has begun to be disrupted by regions of dense material that have formed out in the CGM. These divert the flow, leading to a complex velocity field. Accordingly, the morphology and velocity field of the ensemble of cloud particles is more complex, with outflowing and inflowing regions (that often overlap).

At 0.75 Gyr, the wind in the $\eta_{M}^{\mathrm{cl0}}=5$ simulation (right-hand columns) is already very different to the other simulations. The wind bicone is much narrower and a dense shell of slow moving gas has formed on the edge of the outflow. Much of the CGM is dominated by inflowing material.
There are a large number of cloud particles in the immediate vicinity of the galaxy that have not been efficiently accelerated away. In fact, there are a handful of particles falling inwards close to the disc. The particles that have been accelerated away are limited to the region on the edge of the wind. By 2.25 Gyr, the quantity of dense material out in the CGM has increased, significantly diverting the wind. This also leads to a complex spatial distribution of cloud particles as well as inflowing components. Very close to the disc, there are a mixture of outflowing and inflowing cloud particles.

Turning our attention to the maps of $\sigma_\mathrm{cl}$ in \cref{fig:slice_mvar_153,fig:slice_mvar_460}, it can be seen that the velocity dispersion of cloud particles contained within a single cell can reach over $100\,\mathrm{km\,s^{-1}}$. Thus, on the spatial scale of our gas cell resolution, the motion of the cloud material is not well described by a single valued velocity field. There are two main causes of this non-zero velocity dispersion. Firstly, there are regions where populations of outflowing cloud particles intersect populations of inflowing cloud particles. This can occur out in the CGM (for example, this can be seen at 2.25 Gyr for the $\eta_{M}^{\mathrm{cl0}}=1$ simulation) or close to the galaxy in a low altitude fountain flow (this is seen especially in the $\eta_{M}^{\mathrm{cl0}}=5$ simulation). The second, often more dominant, cause of this velocity dispersion is the differential acceleration of clouds of different masses when exposed to the same background wind. As we demonstrated in \cref{fig:windtunnel}, lower mass clouds are accelerated faster than higher mass clouds. The values of $\sigma_\mathrm{cl}$ of $\sim50 - 120\,\mathrm{km\,s^{-1}}$ seen within the wind (particularly apparent in \cref{fig:slice_mvar_153} for the $\eta_{M}^{\mathrm{cl0}}=1$ simulation, but present in all runs) occur because lower mass clouds overtake higher mass clouds.
Both of these effects mean that the dominant component of the dispersion is in the radial direction for most cells (though a more realistic, less spherically symmetric setup might alter this).

\begin{figure} 
\centering
\includegraphics{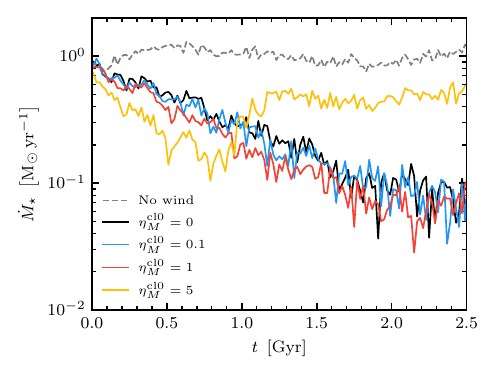}
\caption{SFR as a function of time for simulations without a wind and for different input cloud mass loading factors. We derive the SFR as the mass of stellar material formed within a 20 Myr window. Without a wind, the SFR remains steady. With the inclusion of a wind, inflows from the CGM are suppressed, leading to a reduction in the SFR as the ISM is consumed by star formation. Increasing
the cloud mass loading factor results in further reductions in the SFR as ISM material is ejected from the galaxy. However, for the $\eta_{M}^{\mathrm{cl0}}=5$ simulation, the preventative feedback of the wind becomes ineffective at around 0.8 Gyr, leading to a return to steady-state star formation.}
\label{fig:sfr_mvar} 
\end{figure}

\subsubsection{SFRs, inflow/outflow rates and loading factors}
In \cref{fig:sfr_mvar} we plot the SFR as a function of time for the three simulations presented above, as well as for the simulation with no wind
($\eta_{M}^{\mathrm{w0}}=\eta_{M}^{\mathrm{cl0}}=0$)
and the simulation with a wind but no cloud component ($\eta_{M}^{\mathrm{cl0}}=0$). We derive the SFR from the mass of stellar material formed within 20 Myr windows. Equivalents of the no wind and $\eta_{M}^{\mathrm{cl0}}=0$ simulations have been presented in \citetalias{Smith2024} and we refer the reader to that work for a detailed analysis. In the no wind simulation, the consumption of ISM gas is balanced by inflows from the CGM, resulting in a steady SFR close to $1\ \mathrm{M_\odot\,yr^{-1}}$. With a wind (but no cloud component), the SFR starts off close to the no wind case, but drops steadily, reaching $0.1\ \mathrm{M_\odot\,yr^{-1}}$ by 2 Gyr. As shown in \citetalias{Smith2024}, the reduction of the SFR occurs because the wind suppresses the resupply of gas from the CGM. The wind is therefore a form of preventative feedback. The $\eta_{M}^{\mathrm{cl0}}=0.1$ simulation produces essentially the same SFR evolution, but the $\eta_{M}^{\mathrm{cl0}}=1$ simulation produces a marginally lower SFR. This is because, in addition to the preventative feedback of the wind, the launching of clouds provides an element of ejective feedback, removing star forming gas from the disc. This is strongest in the $\eta_{M}^{\mathrm{cl0}}=5$ simulation. It can be seen that the SFR initially declines much faster than the other simulations. However, around 0.8 Gyr the SFR suddenly increases and then reaches a steady state at around $0.5\ \mathrm{M_\odot\,yr^{-1}}$. This indicates that the preventative feedback of the wind has suddenly become ineffective, as we shall explore when examining the properties of the outflows.

\begin{figure*} 
\centering
\includegraphics{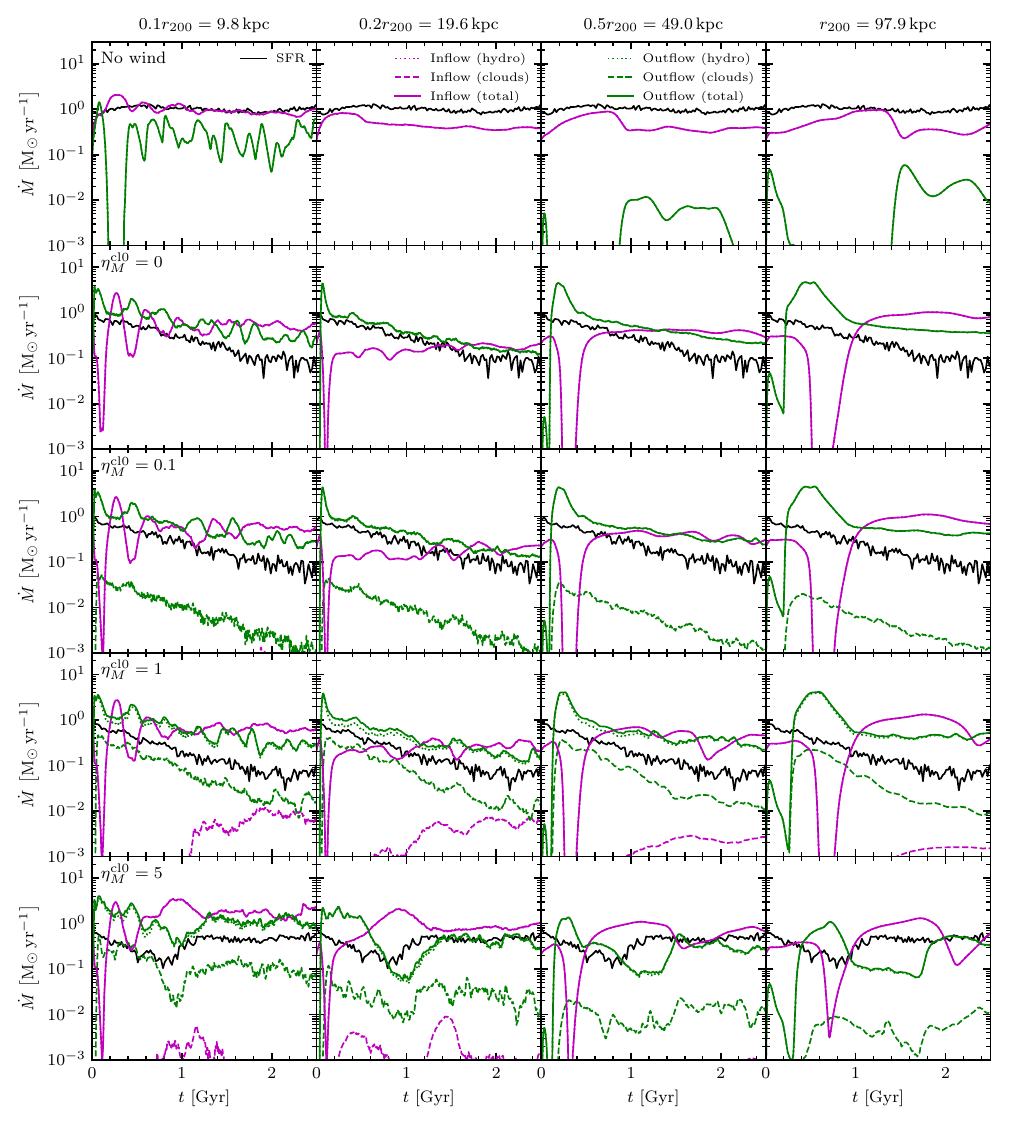}
\caption{SFR, mass inflow and outflow rates through spherical reference surfaces as a function of time. The different columns correspond to measurements taken at 0.1, 0.2, 0.5 and 1 $r_{200}$ (the SFR
is repeated in each column for reference). The locations of these reference surfaces are indicated in \cref{fig:slice_mvar_153,fig:slice_mvar_460}. Each row corresponds to a different simulation, with various input loadings. We plot the total inflow/outflow rates with solid lines and use dotted and dashed lines to indicate the contribution from the resolved hydro and cloud particles, respectively. Without any wind, the SFR remains high, due to CGM inflows.  When a hot wind is included, unless $\eta_M^\mathrm{cl0}$  is very large (bottom row), strong preventive feedback reduces accretion, with the SFR remaining lower than the hydrodynamic outflows through all the measurement surfaces.  At all radii and in all simulations, the cloud mass fluxes remain well below the hydro mass fluxes, even when $\eta_M^\mathrm{cl0} \gg \eta_M^\mathrm{w0}$}.
\label{fig:mflows_mvar} 
\end{figure*}

In \cref{fig:mflows_mvar} we plot the mass inflow and outflow rate through four reference spheres around the galaxy with galactocentric radius $r$ of 0.1, 0.2, 0.5 and 1 $r_{200}$ for the simulations shown in \cref{fig:sfr_mvar}. We also plot the SFR of each simulation for reference. We calculate and plot the mass flow rates for the gas cells (the ``hydro'' component) and cloud particles separately, as well as plotting their sum. For the gas cells,
following \citetalias{Smith2024}, we determine the mass flow rates by considering the fluxes through a sphere at the reference radius. This is achieved by discretizing the reference sphere into $N_\mathrm{pix}$
equal area pixels using the \textsc{HEALPix} library \citep{Gorski2011}. In order to guarantee that the inter-pixel spacing is finer by a factor of a few than the diameter of the smallest cells intercepted by
the sphere, we adopt $N_\mathrm{pix}=786432$. For each pixel centre, we search for the nearest Voronoi mesh-generating point and hence the gas cell within which the pixel is located. Gas cell properties are then mapped on the pixel. The mass flux per unit area through the pixel is then
\begin{equation}
\mathcal{F}_M = \rho v_\mathrm{r},
\end{equation}
where $\rho$ is the cell density and $v_\mathrm{r}$ is the radial velocity (i.e. normal to the spherical reference plane). We can then select only pixels with positive (negative) $v_r$
to compute the mass outflow (inflow) as
\begin{equation}
\dot{M}^{\mathrm{hydro}}_\mathrm{out(in)} = A \sum \mathcal{F}_M,
\end{equation}
where we sum over the selected pixels, each with equal area $A = 4\pi r^2 / N_\mathrm{pix}$. For cloud particles we obtain the mass outflow (inflow) rates as
\begin{equation}
\dot{M}^{\mathrm{cl}}_\mathrm{out(in)} = \frac{\sum m v_\mathrm{r}}{\Delta r},
\end{equation}
where the sum runs over all cloud particles with positive (negative) radial velocity located with in a spherical shell of thickness $\Delta r$. We adopt $\Delta r = $ 2 kpc, 4 kpc, 10 kpc and 20 kpc at $r =$ 0.1, 0.2, 0.5 and 1 $r_{200}$, respectively. In contrast to the cloud particles, all wind particles have already recoupled well inside $0.1r_\mathrm{vir}$.

As shown in \citetalias{Smith2024}, in the simulation with no wind (top row), inflows are completely dominant at all four radii. The inflow rate through 0.1 $r_{200}$ remains constant at about 1 $\mathrm{M_\odot\,yr^{-1}}$ over 2.5 Gyr.
The SFR is essentially identical to this as it is regulated by resupply of material from the CGM to the ISM. Inflows at larger radii are not quite as constant, due to instabilities arising from
our idealised setup. A small outflow can be seen at 0.1 $r_{200}$, but this only arises from disc material crossing the reference sphere. These should be compared against the other simulations
in order to assess the level of ``true'' wind driven mass outflows. A very small outflow can be seen at late times crossing 0.5 and 1 $r_{200}$. This is an outwardly propagating sound wave arising
from the idealised setup and has negligible impact. With the addition of the high specific energy wind component, but no cloud material ($\eta_{M}^{\mathrm{cl0}}=0$, second row), inflows are efficiently suppressed
at all radii. At 0.1 $r_{200}$, the outflow and inflow rates are roughly equal. Further out, outflow rates exceed inflow rates for the first 1 - 1.5 Gyr. The suppression of inflows
acts as a form of preventative feedback, as described above, causing the SFR to drop as the ISM gas supply is used up.
As examined in detail in \citetalias{Smith2024}, the dropping SFR causes a gradual decrease in wind power in absolute terms, resulting in inflows beginning to reassert themselves at
large radii in the latter half of the simulation. It is worth noting that the outflow rate exceeds the SFR at all times and at all measured radii despite the input wind mass loading, $\eta^\mathrm{w0}_M = 0.32$, being less than unity. In other words, the outflow contains more mass than is being ejected from the ISM, indicating entrainment of CGM material; this will be discussed further below.

When cool clouds are introduced, but at a small mass loading ($\eta_{M}^{\mathrm{cl0}}=0.1$, third row), the results are very similar to the $\eta_{M}^{\mathrm{cl0}}=0$ simulation; the cloud phase is subdominant to
the resolved hydrodynamic phase, so has little impact. However, it can be seen that there is outflowing cloud material at all radii, albeit at a much lower level than the rest of the gas. Increasing
the input mass loading of the cloud material to $\eta_{M}^{\mathrm{cl0}}=1$ (fourth row) results in a similar picture. The hot wind continues to efficiently suppress inflows. A larger fraction of the total outward flowing mass flux is now in the form of cloud particles, scaling approximately linearly with the increase in the input mass loading. However, it is still the subdominant component, despite having over three times the input mass loading than the hot wind. This
is because the hot wind is able to entrain significant amounts of CGM material, boosting the total mass of outflowing material well above the injected hot wind mass. 

Further increasing the cloud input mass loading to $\eta_{M}^{\mathrm{cl0}}=5$ (bottom row) results in a change in behaviour. Initially, inflows are suppressed in common with the other simulations that include a wind.
The total mass outflow rate is marginally higher through 0.1 $r_\mathrm{200}$ than the other cases, primarily (as we shall examine later) due to an increase in the amount of stripped cloud material in the wind.
The SFR drops due to preventative feedback, with the addition of a slight additional reduction due to the higher input mass loading causing more ISM material to be ejected. However, from around 0.5 Gyr onwards
the inflow rates surpass the outflow rates in the inner radii, leading to an uptick in the SFR around 0.8~Gyr as the impact of the restored supply of gas to the ISM is felt. For the rest of the simulation,
the wind is not able keep inflow rates below outflow rates at all radii, resulting in the resumption of a steady state SFR, albeit at a reduced level than the no wind case.

\begin{figure*} 
\centering
\includegraphics{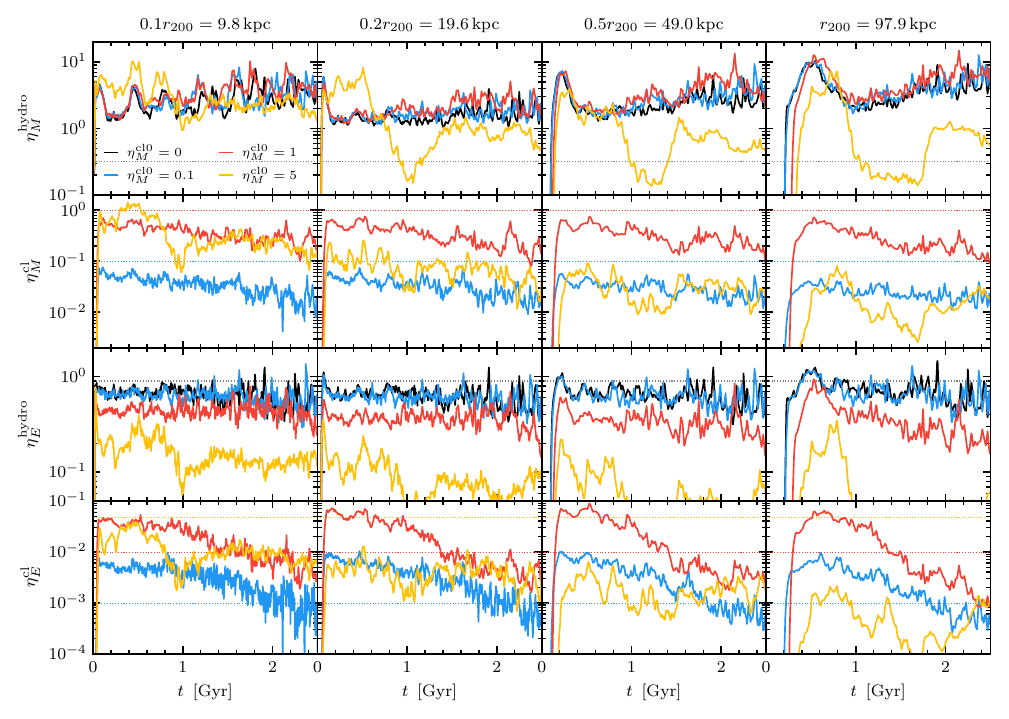}
\caption{Mass loading and energy loadings through spheres
at various radii for simulations with different input cloud mass loadings.
We split the contributions to the loadings into the resolved hydro component (first and third rows)
and cloud particle component (second and fourth rows).
Horizontal dashed lines indicate
the input values (for the energy loadings, this is the combined kinetic and thermal). The simulation without a cool cloud component shows a mass loading that is
around an order of magnitude larger than the input value at all radii, indicating that the outflow
is entraining a substantial amount of CGM gas. The energy loading remains close to the input
value. Adding clouds leads to some additional mass loading in the hydro component at the expense of the cloud
mass loading, as material is stripped. Likewise, the cloud energy loading is increased from the input value
by taking energy from the hydro phase. When the input cloud mass loading is increased to 5,
this results in too much energy being removed from the hot wind to maintain a strong outflow.}
\label{fig:load_mvar} 
\end{figure*}

The relative behaviour of the different simulations can be further understood by examining the emergent mass and energy loadings of the outflows. For both the resolved hydrodynamic and cloud particle
outflows, in an analogous manner
to the definition of the input loadings given in \cref{subsec:launch_and_recouple}, we define the emergent mass loading as the ratio of the instantaneous\footnote{For simplicity, we make the comparison
between the instantaneous outflow rates and the very recent SFR. In order to more carefully examine the link between star formation and emergent outflows, one could attempt to
account for the travel time of the outflow from the galaxy to the reference sphere. However, the outflows lack a single characteristic velocity, which would complicate this procedure.}
 mass outflow
rate through the reference sphere to the SFR averaged over the previous 20~Myr:
\begin{equation}
\eta^{\mathrm{hydro}}_\mathrm{M} = \frac{\dot{M}^{\mathrm{hydro}}_\mathrm{out}}{\dot{M}_\star},
\end{equation}
\begin{equation}
\eta^{\mathrm{cl}}_\mathrm{M} = \frac{\dot{M}^{\mathrm{cl}}_\mathrm{out}}{\dot{M}_\star}.
\end{equation}
We can similarly define energy loadings by examining the ratio of the energy outflow rate to the energy input associated with star formation:
\begin{equation}
\eta^{\mathrm{hydro}}_\mathrm{E} = \frac{\dot{E}^{\mathrm{hydro}}_\mathrm{out}}{u_\star \dot{M}_\star},
\end{equation}
\begin{equation}
\eta^{\mathrm{cl}}_\mathrm{E} = \frac{\dot{E}^{\mathrm{cl}}_\mathrm{out}}{u_\star \dot{M}_\star}.
\end{equation}
The energy outflow rate for the resolved hydro phase is computed in a similar manner to the mass outflow rate. The energy flux through each \textsc{HEALPix} pixel
is
\begin{equation}
\mathcal{F}_E = \rho v_r \left(\frac{1}{2} v^2 + \frac{1}{\gamma - 1}c^2_\mathrm{s} \right),
\end{equation}
for magnitude of the total velocity, $v$, and sound speed, $c_\mathrm{s} = \sqrt{\gamma P / \rho}$. The total energy outflow rate is then
\begin{equation}
\dot{E}^{\mathrm{hydro}}_\mathrm{out} = A \sum \mathcal{F}_E,
\end{equation}
where the sum runs over all pixels with positive $v_r$. The equivalent quantity for the cloud particles is
\begin{equation}
\dot{E}^{\mathrm{cl}}_\mathrm{out} = \frac{\sum m  v_\mathrm{r} \left(\frac{1}{2}v^2 + u \right) }{\Delta r},
\end{equation}
where $u$ is the specific internal energy of the cloud and the sum runs over all cloud particles with positive radial velocity located with in a spherical shell of thickness $\Delta r$.

We plot the mass and energy loadings for the simulations (except the no wind case, which has no appreciable outflows) in \cref{fig:load_mvar}. For the simulation
with no cloud material, the mass loading is between $\sim1 - 8$ at all times and at all radii. Given the input mass loading of the hot wind is only 0.32, this indicates that
a significant amount of inflowing CGM material is being turned around. The energy loading is relatively constant at all radii close to the input value of 0.9, indicating minimal
losses. The $\eta_{M}^{\mathrm{cl0}}=0.1$ and 1 simulations do not substantially differ from the simulation without cloud material with regards to the (hydro) mass loading.
There is a very slight enhancement overall which originates from material stripped from cloud particles.
For both the $\eta_{M}^{\mathrm{cl0}}=0.1$  and 1 cases, the mass loading is a factor $\sim$3 to 10 below the injection value, decreasing in time. 
The missing mass is a result of a combination of shredded clouds (which boosts the hydro mass loading) and clouds which fail to reach the
reference sphere. The latter contribution increases with time as the opening angle of the hot wind decreases (as described above), leading to more cloud particles falling out of the
flow at low altitude or never being entrained to begin with. The input energy loading in the cloud phase is very small, corresponding to the initial launch velocity of $100\ \mathrm{km\,s^{-1}}$
and temperature of $10^4\ \mathrm{K}$; we plot these input loadings with a horizontal dashed line on \cref{fig:load_mvar}. It can be seen that the emergent energy loading
of the clouds, $\eta^{\mathrm{cl}}_\mathrm{E}$, is initially $\sim5$ times larger than the input at all radii. This is due to the efficient acceleration of clouds by the hot wind. The signature of this transfer of energy 
from the high specific energy wind to the clouds can be seen in the emergent energy loadings of the resolved hydro phase, $\eta^{\mathrm{hydro}}_\mathrm{E}$; there is a corresponding deficit relative to the case with no cloud material. $\eta^{\mathrm{cl}}_\mathrm{E}$ decreases with time in a similar manner to $\eta^{\mathrm{cl}}_\mathrm{M}$ as the mass of cool clouds being
entrained in the wind drops.

The simulation with a higher input cloud mass loading of 5 exhibits different behaviour. For the first ${\sim}1$ Gyr of the simulation, the emergent mass loading in the hydro phase is
up to a factor of a few larger than the other simulations at 0.1 $r_{200}$.
Contrastingly, the mass loading in the cloud phase is much lower than the input. As we shall confirm later, this
indicates that the enhancement in the resolved hydro phase comes primarily from shredded cloud material. However, the emergent energy loading in the hydro phase is almost an order of
magnitude lower than the input value. The high specific energy wind loses kinetic energy to the clouds. None the less, it can be seen that the emergent energy loading in the clouds
is actually lower than the input value and does not explain the deficit (as is the case with lower input cloud mass loadings). The hot wind also experiences enhanced radiative cooling 
losses in two forms. Firstly, the increased mass of clouds in the wind results in greater losses via the (subgrid modelled) TRML of clouds. Unlike the kinetic energy transfer, this
energy is not gained by the clouds but is instead lost from the wind. Secondly, the large amounts of mass stripped from the clouds into the hot wind increases
the radiative cooling within the hot phase itself by increasing its density and dropping its temperature towards higher valued regions of the cooling function. Thus, the increased input mass
loading of clouds results in a ``poisoning'' of the wind, leading to a sudden drop in the total mass loading at 0.2 $r_{200}$ and beyond from around 0.5 Gyr onwards.
The wind's capacity to perform preventative feedback is blunted, resulting in the restoration of inflows and subsequent star formation described above.

\begin{figure*} 
\centering
\includegraphics{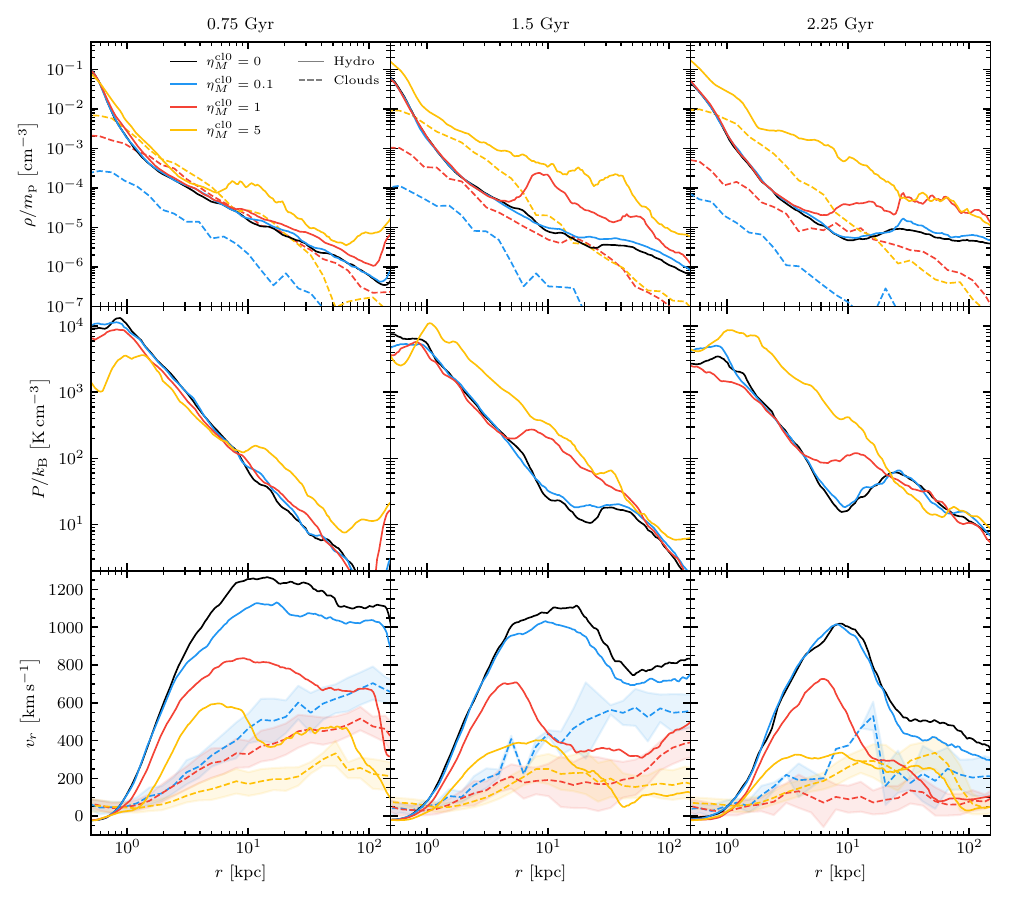}
\caption{Radial profiles of various quantities in the wind region, defined as a truncated cone expanding upwards from the disc with a
cylindrical radius of 2.5~kpc (the disc scale length) in the disc mid-plane and opening angle of 10 degrees (this region is indicated in \cref{fig:slice_mvar_153,fig:slice_mvar_460}).
The top row shows density profiles for hydro (solid) and cloud particle (dashed) components. Note that for the cloud particles, this is the average density of cloud
mass in the region not the internal density of the clouds themselves. The middle row shows pressure profiles for the hydro component, though
due to our assumption of pressure equilibrium this is also the internal pressure of the clouds. The bottom row shows velocity profiles of
hydro (dashed) and cloud (solid) components. For the clouds, the shaded band indicates the 1$\sigma$ scatter.
Increasing the input cloud mass loading tends to reduce the normalisation of the velocity
profiles for both components as the energy of the hot, fast wind is spread around more mass. This becomes less clear at late
times due to the more messy velocity field (see \cref{fig:slice_mvar_460}).}
\label{fig:profiles} 
\end{figure*}
\begin{figure*} 
\centering
\includegraphics{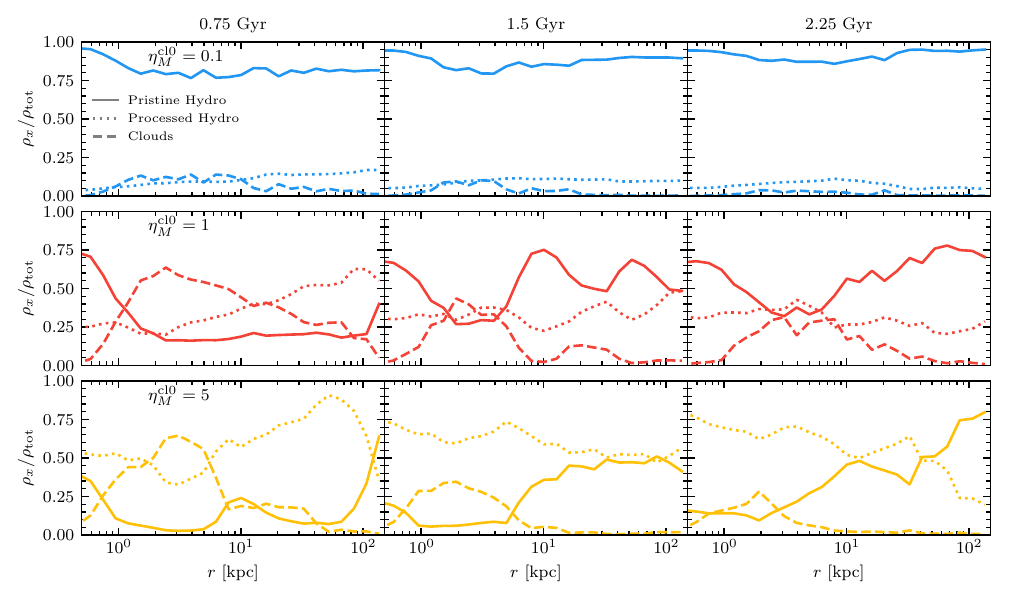}
\caption{Contribution to the total density profiles (i.e. mass fraction) in the wind, shown in \cref{fig:profiles},
by pristine hydro (gas that has never been inside a cloud, solid line), processed hydro (gas that
has been inside a cloud at some point in the past, dotted line) and clouds (dashed line). The columns show three different
times, while the top row, middle and bottom rows show the $\eta^{\mathrm{cl0}}_\mathrm{M}=0.1$, 1 and
5 simulations, respectively. As the input cloud mass loading is increased,
the contribution of the cloud and processed components increases.}
\label{fig:mass_fraction} 
\end{figure*}

\subsubsection{Wind profiles}
In \cref{fig:profiles}, we show various radial profiles of the wind region.
These are calculated in a truncated cone expanding upwards from the disc with a
cylindrical radius of $R=$ 2.5~kpc (equal to the disc scale length) in the disc mid-plane and an opening angle of 10 degrees.
This region is illustrated in the density slices shown in
\cref{fig:slice_mvar_153,fig:slice_mvar_460}. 
For the gas cells, we calculate the profiles in a similar way to the calculation of the
outflow fluxes in order to mitigate fluctuating spatial resolution; at each logarithmically spaced distance from the origin, we discretize the surface into \textsc{HEALPix}
pixels and determine the quantity at the distance as the mean across all the (equal area) pixels.
The profiles are therefore in some sense volume weighted. For the cloud particles, we find all the particles within shells and determine the properties as a mass weighted average.
We plot the simulations at three different times.
In the top row we plot the density profile of the hydrodynamically resolved wind (dashed lines), as well as the mean density of cloud material (i.e. total mass of clouds in a volume, not to be confused with the subgrid internal density of the clouds themselves, solid lines). In the middle row, we plot pressure profiles of the resolved wind. In the bottom row, we plot radial velocities of the wind and clouds. This radial velocity is defined as the velocity parallel to the lines of constant polar angle in the
truncated cone.\footnote{I.e. the recession away from a point 14.2~kpc below/above the disc centre, which is where the apex of the truncated cone would be if it were extended.
If we instead define the radial velocity away from origin (as we do for measuring outflow rates at 0.1~$r_{200}$ and beyond), we get meaningless results at small radii as the gas flow is initially directed vertically out of the disc plane and therefore perpendicular to the vector to the origin for most of the disc. The two definitions of radial velocity converge after a few kpc.}
For the clouds, we also indicate the $1\sigma$ scatter in the cloud velocities with a shaded band.

As shown in \citetalias{Smith2024}, the wind without a cloud component has radial properties close
to the expectations from theories of energy driven winds \citep[e.g.][]{Chevalier1985}, particularly at early times. Away from the ISM, the
density and pressure profiles decline approximately as $r^{-2}$ and $r^{-10/3}$, respectively. The velocity
increases rapidly within 10~kpc as thermal energy is converted to kinetic, before flattening out
at a peak of $\sim1300\ \mathrm{km\,s^{-1}}$.
As described above, later in the simulation, as the wind power decreases due to declining SFR,
inflowing CGM gas begins to impact the wind. Thus, at late times, the density profile flattens
at larger radii and this transition can also be seen in the pressure and velocity profiles.

Adding in cloud material at a mass loading of 0.1 has very little impact on the resolved wind.
At 0.75~Gyr and 1.5~Gyr, the peak velocity of the wind is marginally reduced. The cloud density
profile is approximately an order of magnitude lower than the resolved wind at all radii. At late
times, the uptick in density at large radii seen in the resolved wind is not reflected in the cloud
population. This is because there are not significant inflows of cloud material from outside,
unlike the resolved gas. At 0.75~Gyr, the cloud velocity increases gradually as a function of radius
out to $r_{200}$, indicating a consistent acceleration by the background wind. That said, the clouds
are always much slower than the wind, reaching about half the velocity of the background by $r_{200}$.
At later times, the turnover in the radial profile of the velocity in the wind leads to a
corresponding reduction in the acceleration of the cloud particles.

When the input mass loading of the cloud particles is increased to 1, there is a corresponding
increase in the density profile of cloud material. At 0.75~Gyr, the gas and cloud particle
density profiles have similar amplitudes and slopes throughout much of the halo (we examine
this in more detail below). The density and pressure profiles
of the gas remain similar to the equivalents
in the $\eta_M^\mathrm{cl0}=0$ and 0.1 simulations. However, the wind does not reach such a high
peak velocity, as momentum is transferred to the cloud particles. This also leads to the velocity
profile of the cloud particles being shallower than the $\eta_M^\mathrm{cl0}=0.1$ case.
Eventually, at later times, the slight reduction of the preventative feedback ability of the
hot, fast wind leads
to some re-establishment of inflows, which flattens the density profiles and turns
over the velocity profiles.

Finally, we examine the $\eta_M^\mathrm{cl0}=5$ simulation. At 0.75~Gyr the density
of both the gas and cloud material are higher than the other simulations. The acceleration
of the wind is blunted even further, reaching a peak velocity less than half that of the
simulation without cloud material. Likewise, the cloud particles are also accelerated
more slowly. This reduction in wind velocities, along with the resulting re-establishment of
inflows (as discussed above) leads to a significant enhancement of the density
in both phases at late times. It can be seen in \cref{fig:slice_mvar_460}, however,
that there are significant structures in both the density and velocity
slices which makes a spatially averaged profiles an oversimplification.

\begin{figure*} 
\centering
\includegraphics{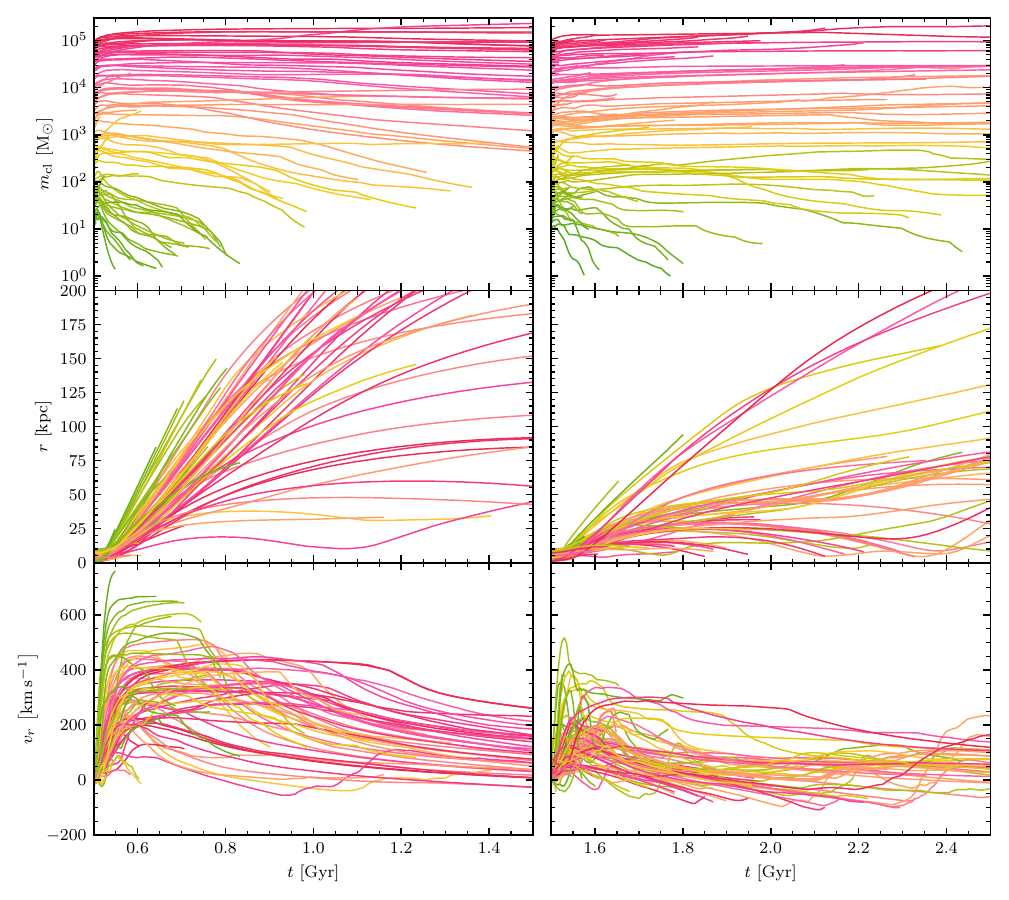}
\caption{Time evolution of cloud mass, galactocentric radius and radial
velocity for a selection of cloud particles created in the 5~Myr preceding 0.5 and 1.5~Gyr
in the $\eta_M^\mathrm{cl0}=1$ simulation.
For the 0.5~Gyr selection (left) we plot half of the particles created
in that time window (selected randomly) to avoid overcrowding the plot.
For the 1.5~Gyr sample, we plot all tracks. Lines are
coloured by their initial cloud mass. Clouds that are initially
low mass are accelerated rapidly, but tend to be shredded.
More initially massive clouds maintain or grow their mass,
but are accelerated relatively slowly. Some clouds
leave the hot wind flow and fall back towards the galaxy.
}
\label{fig:ump_track} 
\end{figure*}

In \cref{fig:mass_fraction}, we plot the relative mass fraction of cloud and wind material
in the outflows as a function of radius, using the same spatial cut as the radial profiles
shown in \cref{fig:profiles}. We consider mass in cloud particles, ``pristine'' gas, that
has never been inside a cloud particle, and ``processed'' gas, that has been inside
a cloud particle at some point in the past. To enable the quantification of the pristine and
processed mass fractions, we make use of a passive scalar ``dye''. The mass fraction
of the dye is initialised to zero in all gas at the beginning of the simulation.
Mass that is transferred from a cloud particle to a gas cell is assigned a dye
mass fraction of unity. The dye is advected by the hydro scheme.
It can be seen that for the simulation with an input cloud mass loading of 0.1 (top row) the
mass in the outflow region is dominated by pristine gas at all times and distances.
For the $\eta^\mathrm{cl0}_M=1$ simulation (middle row) at 0.75~Gyr, once outside of the ISM,
the cloud particles are the dominant mass component until $\sim$10~kpc, with a
peak at 2~kpc, followed by a decline. This decline is matched by an increase
as function of radius of the processed gas component. This occurs as the
clouds are gradually shredded into the wind as they travel. At later times,
the reduction in the absolute outflow rates and the increase in
inflows leads to the pristine gas component becoming more dominant.
The $\eta^\mathrm{cl0}_M=5$ case (bottom row) has a similar, but more extreme behaviour.
As discussed previously, the wind struggles to propel such a large mass
of clouds to large distances. Thus, at 0.75~Gyr the clouds only dominate
over the other mass components inside 0.1~$r_{200}$. However, the result
is the wind is then overwhelmingly dominated by material stripped from
clouds. This is true even at later times when inflows begin to become
important.

\begin{figure*} 
\centering
\includegraphics{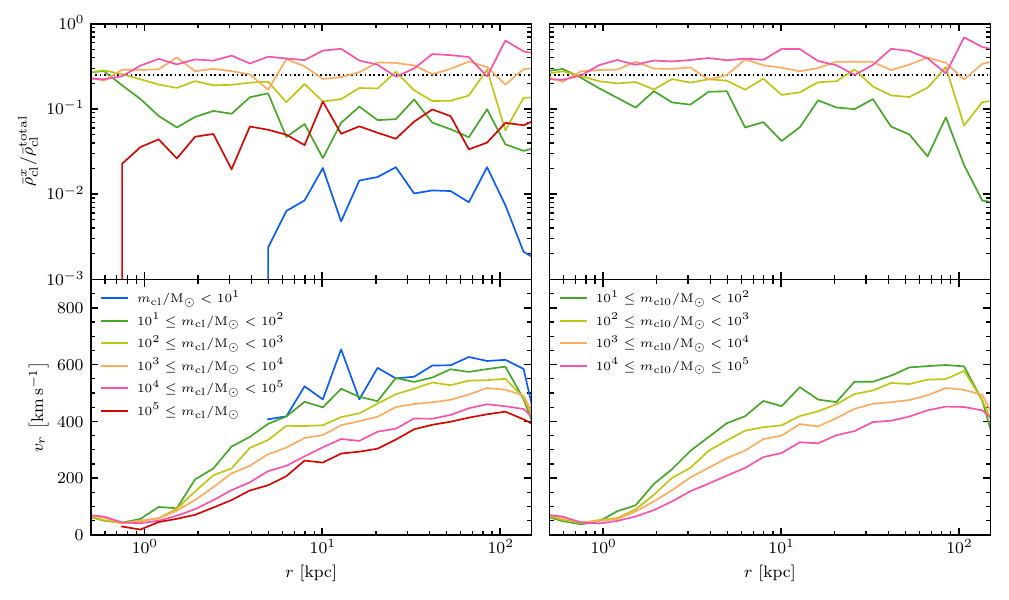}
\caption{Radial profiles at 0.75~Gyr in the $\eta_M^\mathrm{cl0}=1$
simulation within the wind region (defined as in \cref{fig:profiles})
logarithmically binned by current cloud mass (left) and initial cloud mass (right).
In the top row, we plot the relative contribution of clouds in a given
bin to the total cloud mass density. 
The unmodified input cloud mass function ($dN/dm \propto m^{-2}$)
would result in an equal contribution from the four logarithmic bins
in the range $10 - 10^5\ \Msun$; this value is indicated for
reference with the horizontal dashed line.
The initial mass distribution
of clouds is modified with increasing radii as low mass clouds
shred and higher mass clouds grow. In the bottom row,
we plot mass weighted radial velocity profiles for each
cloud mass bin. Lower mass clouds are accelerated much
faster than higher mass clouds, giving rise to
a difference of around 200~$\kms$ between the lowest
and highest mass clouds at a given location.
}
\label{fig:ump_pop_profiles} 
\end{figure*}

\subsubsection{Cloud evolution}
In the preceding figures, we have examined the properties of the entire population
of clouds in the simulation. However, the Lagrangian nature of our
approach to modelling the clouds enables us to follow the evolution
of individual clouds over time, as we demonstrated in
\cref{fig:windtunnel} for our simple wind tunnel tests.
In \cref{fig:ump_track}, we plot
some examples of evolutionary tracks for clouds extracted from
the $\eta^\mathrm{cl}_M=1$ simulation. At 0.5~Gyr and 1.5~Gyr, we
find all of the cloud particles created in the preceding 5~Myr.
We plot their cloud mass, galactocentric distance and radial velocity as a function
of time. For the 0.5~Gyr sample, we only plot half of the tracks (selected randomly)
to avoid overcrowding the plot, while for the 1.5~Gyr sample we plot all tracks.\footnote{Note that the combination of our adopted initial
cloud mass function ($\mathrm{d}N/\mathrm{d}m \propto m^{-2}$) and our 
choice to use a single initial cloud particle mass leads to the tracks initially
being uniformly distributed in the logarithm of the initial cloud mass.}
The colour used to plot the tracks reflect the \textit{initial} cloud mass.

There are general trends that are qualitatively consistent with the results of our
simple wind tunnel test. The clouds with the lowest initial masses are accelerated
fastest but often experience a net mass loss at all times. More massive clouds
experience a smaller acceleration but are more likely to grow or keep a stable
mass. However, the spatially and temporally evolving background wind in these
simulations produces additional complexities that are not present in a uniform
wind tunnel. 
As demonstrated above, in the period spanned by our first set of tracks (0.5 - 1.5~Gyr),
the wind structure is more well behaved than during the second period (1.5 - 2.5~Gyr),
in the sense that there are fewer substructures and inflowing features.
In the first set of tracks, it can be seen that clouds with initial masses
below $\sim100\ \mathrm{M_\odot}$ are rapidly accelerated up to the wind velocity
(see \cref{fig:profiles}), at which point their velocity track suddenly flattens out.
The scatter in the peak velocity reached is predominantly a result of the
variation of the wind velocity with polar angle; the flow is faster closer to
the centre of the wind bicone than at the edges. This can be seen
in \cref{fig:slice_mvar_153}, while \citetalias{Smith2024} examines this in
more detail.
All of these clouds have been destroyed (in our implementation,
they have lost more than 90\% of their initial mass and are then fully
recoupled to the gas) within 250~Myr. Many are destroyed on much shorter
time-scales. Several of the cloud particles reach or exceed $r_{200}$
(97.9~kpc). The wind velocity drops suddenly in this region at these times,
leading to a sudden deceleration of the clouds. A few of the low mass
tracks show an earlier and more gentle deceleration; these are particles
that are close to the edge of the wind and stray into regions of
low wind velocity. All of these are destroyed before they fall out
of the outflow region. For more massive initial clouds, we see
similar trends, but the evolution is slower. Clouds experience
lower accelerations, so they are not as closely coupled to the
background gas. There is a large scatter in the tracks even
with clouds of similar initial masses, seeded by variations
in the local properties of the background wind and then amplified
with time. Some clouds grow at all times, others lose mass at later times.

For the particles selected at 1.5~Gyr, the general trends remain,
but more complex trajectories are evident. The properties of the
background wind fluctuate
significantly in space and time. The background
velocity field experienced by a cloud can change suddenly
as the opening angle of the wind changes as a function of radius
due to collimation by the CGM. It can also change due to
fluctuations originating upstream (caused, for example, by the changing SFR or substructures
close to the disc) catching up with a cloud particle
moving significantly slower than the wind.
The lowest mass clouds are
closely coupled to the background flow, but the lower accelerations
experienced by more massive clouds mean that they are slow to
react to these changes. 
This leads to an enhanced fraction of clouds falling out of the flow.
Many clouds turn around and
accelerate back towards the galaxy. 
These can be seen in \cref{fig:slice_mvar_460}, flowing
down the edges of the outflow.
Some fall back into the ISM,
their density contrast drops below our threshold value of 10 and
are recoupled. Others re-enter outflows as they
near the galaxy and are accelerated away from the galaxy once again.
The more massive a cloud, the more likely it is to survive
by growing in response to the sudden increase of relative
velocity as it enters the outflow. On the other hand,
if it is too massive, it takes too long to arrest its fall
before it reaches the ISM.

In \cref{fig:ump_pop_profiles}, we once again show radial profiles 
(in the outflow region, defined as in \cref{fig:profiles})
of the cloud
population from the $\eta_M^\mathrm{cl0}=1$ simulation at 0.75~Gyr, but
binned by cloud mass. In the left panels we split the population
by the current cloud mass while in the right panels we split by the
initial cloud mass. We show the contribution of a given mass bin to
the total cloud density as well as the mass weighted mean radial
velocity for each mass bin. Our mass bins are logarithmically
spaced. In combination with our adopted initial cloud mass function
of $\mathrm{d}N/\mathrm{d}m \propto m^{-2}$, this means that
the mass bins within our initial cloud mass range ($10 - 10^5\ \mathrm{M_\odot}$)
contribute equally to the total cloud mass at small distances from
the galaxy. However, as lower mass clouds are shredded while
larger mass clouds remain stable or grow (as shown previously),
this means that the cloud mass at larger radii is dominated by
the more massive clouds (or, indeed, the clouds that started
off as the most massive). Turning to the radial velocity
profiles, it can be seen that there is a wide spread
in normalisations of the profiles. As previously shown,
the more massive the cloud, the lower acceleration it experiences.
This leads to a spread of around 200~$\mathrm{km\,s^{-1}}$ between
the least and most massive bins. This is a major contributor
to the cloud-to-cloud velocity dispersions within cells
shown in \cref{fig:slice_mvar_153,fig:slice_mvar_460}, as
more massive clouds are overtaken by less massive clouds.

\begin{figure*} 
\centering
\includegraphics{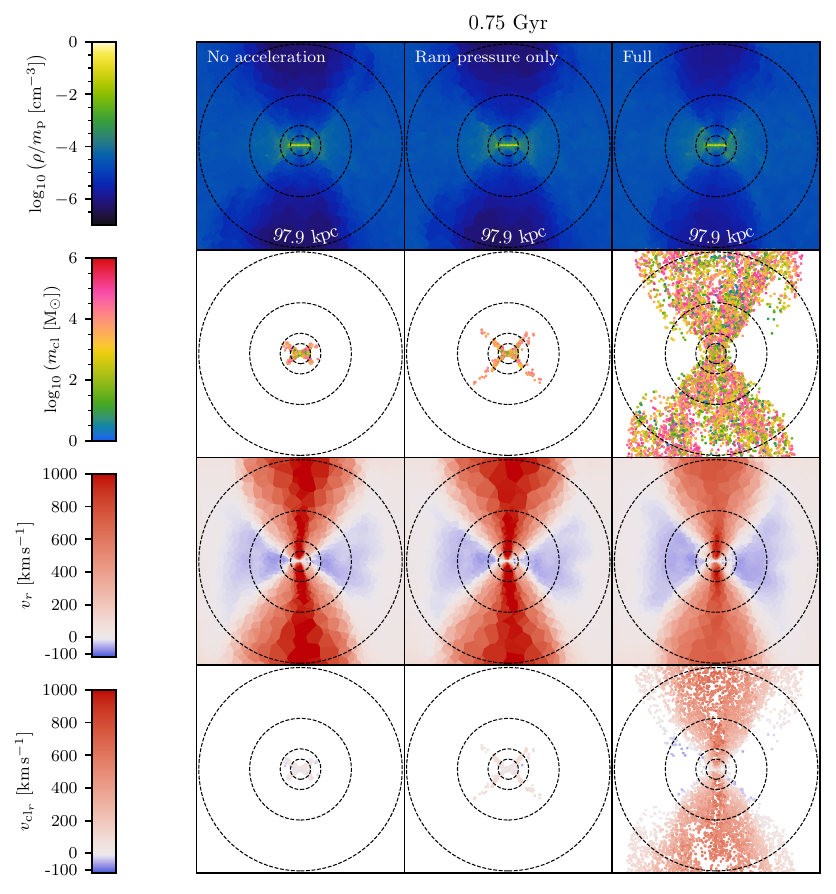}
\caption{Slices as in \cref{fig:slice_mvar_153} for simulations with
$\eta_M^\mathrm{cl0}=1$ at 0.75~Gyr. In the simulation shown
in the left column, we disable both cloud growth and acceleration
by ram pressure. In the middle column, we disable cloud growth
but allow acceleration by ram pressure. The right
column contains the fiducial physics simulation. Without
a non-zero mass growth term (which also brings
momentum into the cloud), clouds are not efficiently
accelerated away from the galaxy and are rapidly destroyed.
}
\label{fig:slice_mvar_noacc} 
\end{figure*}

\subsubsection{Ram pressure is not the dominant source of acceleration}
Finally, in order to demonstrate that it is the accretion of high momentum
material into the clouds that provides their high velocities, we
perform two resimulations of the $\eta_M^\mathrm{cl0}=1$ case.
We set the cloud mass growth rate, $\dot{m}_\mathrm{grow}$, to zero.
In the first simulation, we also set the drag coefficient, which
appears in \cref{eq:pdrag}, to zero. This means that the clouds
will travel ballistically after being launched
with the initial velocity of $100~\mathrm{km\,s^{-1}}$,
only experiencing acceleration due to
gravity. They still lose mass due to shredding, however. In the
second re-simulation, we set the drag coefficient back to our
fiducial choice of 1/2 to see if this source of acceleration is
enough to propel clouds away from the galaxy.
In \cref{fig:slice_mvar_noacc}, we show slices (equivalent to
\cref{fig:slice_mvar_153}) of these simulations after
0.75~Gyr, alongside the fiducial full model run.
It can be seen that only the fiducial simulation
including cloud mass growth and associated momentum
gain from the hot, fast wind leads to clouds being
accelerated throughout the halo.
The two re-simulations only show evidence of a small
number of clouds in an x-shaped distribution (as seen in the slice).
These are the only clouds that have been able to survive.
As a reminder, in this work, 
we give the cloud particles an initial vertical kick of 
$100\ \mathrm{km\,s^{-1}}$ away from their birth site.
However, this is added to the initial velocity
of the star forming gas from which they were born, so
the cloud trajectory is not exactly vertical. Without
the mass growth and acceleration terms, cloud particles
that end up in the hot outflow are rapidly destroyed
before they can be accelerated by the drag term (in
the simulation that includes it). Only those cloud
particles launched exactly along the edge of the
wind bicone (where the relative velocity is smallest)
manage to survive, giving rise
to the x-shaped distribution in the slice. Even then,
these are only the most massive clouds. Lacking
an efficient acceleration source, this very small
amount of cloud mass quickly turns around
and fall back to the galaxy. 
This confirms, for the full model, that the acceleration
due to momentum-rich mass being
accreted into the cloud (the first term
of \cref{eq:pdot}) dominates over the
drag term and that our initial
kick of $100\ \mathrm{km\,s^{-1}}$
is negligible compared to the velocities
the clouds reach when accelerated by the hot wind.

\section{Discussion} \label{sec:discussion}
\subsection{Model applications, interpretation and extensions}
In this work, we have demonstrated \textsc{Arkenstone}
in idealised non-cosmological tests in order to highlight the main
features of our scheme in a clean manner.
In practice, the model
is intended to be deployed in large-scale cosmological
simulations. Bennett et al. in prep. will present
the first application of the \textsc{Arkenstone-Hot} part of the scheme (introduced in
\citetalias{Smith2024}) in cosmological simulations.
The parts of the model presented in this article
will be included in a cosmological setting in another future work.
None the less, we stress that because the
cloud--wind interaction terms depend solely on
local properties (i.e. without reference to
larger scale galaxy or halo properties), this aspect of the
scheme is already fully deployable in cosmological simulations.

The highly complex small-scale
interactions of stellar feedback with the ISM that
give rise to galactic winds are intentionally
omitted in our model (via the use of initially
hydro decoupled particles) because it is impossible
to capture them properly at the resolution accessible
in large-scale cosmological simulations. Attempting to
launch galactic winds by direct injection
of feedback energy into the ISM at these resolutions
provides no predictive power. In fact, such an approach actively
hinders the interpretation of results because
the necessarily unphysical ISM behaviour
and emergent wind properties
are likely to be over-interpreted.
Instead, in our scheme, the input
mass and energy loading, along with the initial
cloud mass function, encodes the properties of the
wind as it leaves the ISM, distilling all of the
unresolvable small-scale physics into
an interpretable and
numerically robust parametrisation.

In this work, we used fixed values of mass and energy loadings
for both wind phases for simplicity. The values were chosen
to demonstrate various regimes of cloud--wind co-evolution.
In practice, as discussed in \citetalias{Smith2024},
it is more appropriate for the loading factors to be
adaptively selected during the simulation. There are
essentially two approaches that can be adopted.
Firstly, the loadings can be scaled with
properties of the dark matter halo
\citep[e.g. local dark matter velocity dispersion, as in][]{Oppenheimer2008,Okamoto2010}
or the galaxy \citep[e.g. stellar mass, as in][]{Dave2016}.
These scalings can be tuned to match some observational constraints on
galaxy properties \citep[e.g.][]{Vogelsberger2013,Pillepich2018} or
to capture a link between galaxy and wind properties seen in higher
resolution simulations \citep[e.g.][]{Dave2016,Dave2019}.
Scaling with large-scale properties permits the placing
of a wind with particular properties into a particular galaxy/halo,
enabling controlled experiments to be performed or simply the
recovery of certain galaxy properties (in the case of tuned scalings).
An alternative approach is to scale the loadings with
local ISM properties based on very high resolution
simulations that are more able to resolve the relevant physics, providing more predictive power.
One can go one step further and allow wind/cloud particles to have
their initial velocity and temperature drawn from a distribution
\citep[see e.g.][]{Kim2020a} rather than being single valued.
This means that the winds (and their impact) are also more
sensitive to other aspects of the entire galaxy formation
model (the ISM model in particular), which can be an
advantage or disadvantage depending on the aims of
a given application. Therefore, the \textsc{Arkenstone}
implementation is compatible with either method of scaling
the loadings.

In this work, we only tested one model of cloud--wind
interaction.
However, the underlying \textsc{Arkenstone} methodology permits
the \citetalias{Fielding2022} model to be easily amended or
entirely replaced. As described in \cref{subsec:cloud_particle_ev},
all that is needed is to provide a functional form of the mass, momentum, energy
and metal exchange rates between the cloud and background
wind. Physics not included in this work, for example the impact of
thermal conduction and magnetic fields,
can be included by changing how these various exchange terms
arise from the current properties of the clouds and the local
ambient wind.
Likewise, a more detailed treatment of the internal
phase structure of the cloud (the division between
ionised, neutral and molecular gas)
could be adopted with a suitable analytic
prescription \citep[e.g.][]{Vijayan2024}.
Processes leading to cloud splitting
can be included by relaxing the assumption that the cloud number,
$N_\mathrm{cl}$, is an immutable property of the cloud
particle.\footnote{The only constraint is that cloud particles
can only contain a population of identical clouds because clouds
of different masses follow different trajectories, precluding
our approximation of the population as a point mass.
For example, one could split all clouds in a particle into
two equal clouds, resulting in a doubling of $N_\mathrm{cl}$.
However, any uneven splitting of clouds requires a splitting
of the cloud particle itself into two (or more) particles
that each contain homogeneous cloud populations. This functionality
can be added in the future if needed.} Furthermore,
any additional physics that impacts the properties of
the wind as it leaves the ISM can be folded into the
determination of the input mass and energy loadings.
As an example, we can consider the potential impact of
cosmic rays. Cosmic rays may increase the
mass and energy loadings of cool gas leaving the
ISM \citep[see e.g.][]{Rathjen2023,Armillotta2024}. This effect
can be included by altering the input mass and energy
loadings, ideally by measuring them in a high resolution
simulation. If one wished to include the effects
of cosmic rays on the acceleration of clouds
further away from the ISM, this could be achieved
by modifying the cloud--wind interaction terms
to account for this by providing an additional
acceleration term (with an appropriate backreaction).

While we have primarily discussed
the \textsc{Arkenstone} cloud model in the context of
stellar feedback driven winds, the scheme can be used
to model the behaviour of unresolvable cool clouds in other contexts.
For example, the scheme could be used to model the evolution of cool
material entrained in AGN outflows. Furthermore,
the creation of clouds is not limited to ejection in winds.
With the inclusion of suitable cloud creation criteria,
\textsc{Arkenstone} could be used to model the formation
of clouds via thermal instability in the CGM/IGM and their
subsequent precipitation \citep[see e.g.][]{Field1965,McCourt2012,Sharma2012,Voit2015}.

Finally, we emphasise that a key motivation behind the \textsc{Arkenstone} model is to enable us to make predictions of the observational signature of cool material in winds and the CGM and, importantly, its dependence on the assumed small-scale physics.
Our model makes
predictions for both the quantity, spatial distribution and small-scale properties
of the clouds. We intend to translate these into mock observations of the clouds in absorption and emission. We leave a detailed discussion of this topic for a future work, barring a few brief remarks.
As explained in \cref{sec:methods}, cloud particles represent the distribution of cloud material in a statistical sense; using one particle for one cloud would require a prohibitively expensive number of resolution elements. The particles are point masses with no extent, so one cannot compute an absorption spectra by simply ray-tracing through the gas distribution.
\cite{Hummels2023} present a tool, \textsc{CloudFlex}, that makes predictions for absorption-line signatures of cool material in the CGM distributed into complex structures of clouds. \textsc{CloudFlex} places clouds in a CGM in a Monte Carlo fashion, taking as input various free parameters governing the distributions of cloud positions and properties.
Once it has been populated with clouds,
sightlines can be sent through the CGM and absorption spectra can be computed based on the clouds intersected. A similar approach could be used to forward-model \textsc{Arkenstone} outputs, but with the distribution of clouds constrained by the cloud particle locations.
Thus we can make predictions for the observational signature of ``low ions'' (e.g. \ion{C}{II}, \ion{C}{III}, \ion{Si}{II}, \ion{Si}{III}, \ion{Mg}{II}) and \ion{H}{I}. In addition to absorption
from the cool material within
the cloud,
we expect absorption from
higher ions (e.g. \ion{Si}{IV},
\ion{N}{V}, \ion{O}{VI}) within
the intermediate temperature ($\sim10^5\,\mathrm{K}$) gas of the
TRML. This will require
a model for the temperature and
ionisation structure within
the TRML \citep[see e.g.][]{Ji2019,Tan2021a,Chen2023}
which is not a direct
prediction of \textsc{Arkenstone}.
The TRML radiates
energy away as material cools
onto the cloud, primarily in Ly~$\alpha$ emission.
As we know $\dot{m}_\mathrm{grow}$ for our
cloud particles, we can relate this to a
total luminosity \citep[see e.g.][]{Gronke2022}
which is is a good proxy for Ly~$\alpha$ luminosity.
As with the TRML absorption signature,
predicting the emissivity of
metal lines requires a more detailed
model of the TRML \citep[see e.g.][]{Tan2021a,Chen2023}.

\subsection{Comparison to other schemes} \label{subsec:comparison}
\citet{Huang2020} present a subgrid model, named \textsc{PhEW},
in the meshless finite mass (MFM) code \textsc{Gizmo} \citep{Hopkins2015},
where wind particles are treated as collections of clouds in
an analogous manner to our scheme. Their model is inspired
by the results of cloud crushing simulations with
thermal conduction \citep{Bruggen2016}. In this framework,
there is a competition between the tendency of thermal conduction
to lead to the evaporation of clouds versus its ability to extend
the lives of clouds by suppressing instabilities (e.g. Kelvin-Helmholtz)
that would otherwise disrupt the cloud. The model does not include
the effects of radiative cooling onto the cloud (as in the \citetalias{Fielding2022})
model, so there is no possibility for clouds to gain mass or momentum
from the ambient wind and the only form of acceleration is ram pressure.
Particles can only deposit mass into background gas,
rather than needing to handle simultaneous deposition and accretion
of material as in our case. \citet{Huang2022} present an application of the \textsc{PhEW} model
in cosmological simulations. They launch all galactic wind material in
the form of cloud particles.
In other words, \textsc{PhEW} does not treat winds as multiphase,
but as single phase outflows composed entirely of cool clouds.
Unlike our model, there is no background hot, fast wind
that can accelerate cool clouds once they leave the ISM.
Cloud particles are therefore ejected from the ISM (with a brief
hydro decoupled phase, as in our scheme)
and are then decelerated by ram pressure as they impact the CGM, with the cloud
evolution model determining how far they can travel before being
completely destroyed.

The impact of \textsc{PhEW} is compared to
simulations where the wind particles deposit all of their
mass into local gas once they meet the recoupling criteria.
There are two primary impacts of the model. Firstly,
the simulations with the \textsc{PhEW} model are in
better agreement with low-redshift galactic stellar mass
functions (GSMFs) for $M_\star < 10^{11}\ \Msun$ due to increased
wind recycling. This increase in recycling arises
because the \textsc{PhEW} model permits
a more gentle and spatially extended deposition of wind
particle material, rather than a single point injection.
However, we note that schemes similar to the non-\textsc{PhEW}
model \citep[e.g.][]{Pillepich2018,Dave2019} can also achieve good
agreement with the low mass end of the GSMF. The other
major conclusion is that the \textsc{PhEW} model leads
to a very different distribution of metals within CGMs.
However, \textsc{Gizmo} is a completely Lagrangian
scheme in the sense that there are no mass fluxes between
resolution elements. This means that, in the absence
of an additional metal mixing scheme
\citep[for discussions of \textsc{Gizmo}-specific implementations see e.g.][]{Su2017,Rennehan2021}
there can be no metal fluxes between resolution elements.
\citet{Huang2022} report that this drives the differences
between the non-\textsc{PhEW} and \textsc{PhEW} simulations. In both
cases, metals stay locked up in the gas particles into which they are
deposited; the \textsc{PhEW} model spreads out this injection of
metals over many more particles.
It is therefore unclear whether this difference has physical significance or is compensating for a lack of metal mixing in the underlying scheme.

Both \textsc{Arkenstone} and \textsc{PhEW} model the cloud
component of the fluid in a Lagrangian manner. However, an
alternative approach is to use an Eulerian discretisation, known
as a ``two fluid'' or ``multifluid'' approach.
\cite{Weinberger2023} introduce a generalised scheme 
along these lines in \textsc{Arepo},
based on the stratified flow model of \cite{Chang2007}.
Each finite volume resolution element represents a mixture
of two different fluid phases, each with their own state vector.
One can assume each fluid behaves in a similar manner to a
single fluid, except with the addition of terms governing
the exchange of mass, momentum and energy. In practice,
this means solving three Riemann problems at each interface: the
interactions between the first fluid on each side of the interface,
between the second fluid and between the first and second fluid.
Then, additional source/sink terms can be included to capture
any additional interaction physics (such as a cloud--wind interaction)
in an analogous manner to our implementation.
However, in the case of cool clouds embedded in a wind,
cloud -- cloud collisions should not occur, due to
their low volume filling fractions. For this reason, the
two-fluid scheme of \cite{Butsky2024}, implemented in the
adaptive mesh refinement (AMR) code \textsc{Enzo},
models
the cool component as being pressureless (although the
clouds themselves have an internal pressure), which
also has the advantage of simplifying the
problem from the generalised case presented in \cite{Weinberger2023}.
A benefit of multifluid schemes is that they do not require
a second set of resolution elements (though the memory and
computational requirements per element still increases with each additional fluid)
and the fractional volume density of each component is well
defined throughout the domain. Unlike our cloud particle based approach,
multifluid schemes do not have to perform
neighbour searches to associate the cool material with its neighbouring
hot gas. That said, in \textsc{Arkenstone} these searches only take
up $\sim$2\% of the total computational expense.
\cite{Weinberger2023} demonstrate a compelling application
for the multifluid approach, replacing the
\citetalias{Springel2003} eEoS model for the ISM, which assumes a two-phase
medium in each cell with the same velocity, with two fluids that can have relative motion
(but using the same source/sink terms). The general behaviour of the
eEoS and multifluid simulations (idealised and cosmological) are the same
on large scales, but the multifluid scheme allows for the production
of a thin disc of cool material and a thicker disc of warm material,
as might be intuitively expected.

However, in the context of cloud--wind interactions,
multifluid schemes have several drawbacks which
limit their applicability for modeling galactic outflows.
Firstly, each phase in the resolution
element can have its own independent velocity, but the velocity field
for a given phase is single valued. As we demonstrated
in this work, the velocity dispersion between cool clouds on our resolved
scale is typically non-zero. Clouds with different properties
experience differential acceleration even when exposed to the same
background wind, leading to smaller clouds overtaking more massive clouds.
Likewise, we demonstrate fountain flow behaviour with inflowing
clouds passing outflowing clouds. In a multifluid scheme, regardless of
whether the cool phase is modelled as a pressureless fluid or not,
such behaviour is not possible. Intersecting flows of clouds ``collide''
and result in a single, physically meaningless bulk velocity.
This can only be avoided if the spatial resolution in the simulation is
high enough to resolve the inter-cloud separation, in which case
the advantages of the multifluid scheme are lost. \textsc{Arkenstone} occupies the opposite limit;
the cloud particles are collisionless
with respect to each other. The cloud velocity field
can therefore be multivalued on any spatial scale. As currently implemented, our scheme is inappropriate for regimes where collisions between clouds are likely (e.g. the ISM), but this is not a concern for our intended applications.

Secondly, it is difficult to track the time evolution of properties of distinct cloud populations with a multifluid scheme.
Only properties averaged across the volume element are known.
For example, one can know the total mass of clouds in a cell
but cloud scale properties (such as cloud masses or equivalently
cloud radii) are not explicitly tracked. If a scheme could be constructed
such that these properties
were advected around with the fluid flow they would
become poorly-defined when the cool fluid mixes with itself.
For example, if there was a flux of clouds with a particular
mass into a cell that contained clouds of another mass, the information
of the two distinct cloud populations is necessarily lost.
One could use an additional fluid for every
possible configuration of cloud but this would rapidly become numerically
intractable. Thus, cloud models have to make the assumption
that either all clouds everywhere in the simulation domain are identical
or that a cloud's properties can be instantly inferred from the local
properties of a cell, discarding all prior evolution.
One could imagine assuming that the cloud masses in a cell
follow some statistical distribution (either assumed to be universal
or determined from local properties), but this then falls foul of the
single valued velocity problem described in the previous paragraph;
clouds with different masses follow different trajectories, so cannot
be represented by a single fluid.
Additionally, we have shown that the cloud distribution
at a given point in the wind (e.g. \cref{fig:ump_pop_profiles})
emerges from the intersection of the trajectories of different
clouds, launched at different times,
that may have been stripped or grown on their way to that
location, making an estimate of the distribution from purely
local and instantaneous properties difficult.

Finally, we note a potential numerical issue with the multifluid
scheme that could arise in our context. \cite{Weinberger2023} report
that with a fixed mesh their scheme is very diffusive, due to numerical mixing
of the phases from advection errors. Once \textsc{Arepo} is allowed to
operate in its usual mode, with the mesh generating points moving with
the local fluid velocity, these errors are reduced to
a completely negligible level because the fluxes between the cells
are significantly reduced.
In the idealised test
they use to demonstrate this behaviour, the two fluids have the same velocity.
However, in the general case where there is a non-zero relative velocity
between the two fluids, it is not possible to choose a mesh-generating point
velocity
such that both fluids are at rest with respect to the mesh. One could choose to
move the mesh with the velocity of either fluid or the (weighted) bulk velocity
of the combination, but it will lead to a relative mesh -- fluid motion in either one or both
of the fluids. When the relative velocity of both of the phases is small,
the increase in diffusivity may be negligible.
However, as we demonstrate in this work (e.g \cref{fig:profiles}), the relative
velocity between the wind and the clouds can be on the order of $600-800\ \kms$
(or equivalently, $\mathcal{M}\sim6-8$ in the cloud phase). Thus,
any choice for the motion of the mesh would necessarily result
in a large relative mesh -- fluid velocity which may promote advection errors.
In a static mesh code (such as AMR), the choice is not available.

In addition to describing a general framework for a two fluid scheme,
\citet{Butsky2024} also implement a cloud evolution model inspired by
\citetalias{Fielding2022}, so it is instructive to make a direct comparison
to our implementation. As described above, the choice to use a
two fluid scheme removes the ability to take the history
of a cloud in account when determining its future evolution.
\citet{Butsky2024} set the cloud radius based on
instantaneous and local properties of the cell, inspired by the
mist model of \citet{McCourt2018}. The implication of this is
that as subgrid clouds
gain mass they instantly fragment, increasing the number of clouds
in the cell, in order to maintain 
a constant cloud radius. Likewise, as clouds lose mass, they
must be instantaneously merging with other clouds, reducing the cloud number,
in order to preserve the mandated cloud radius.
While it may be appropriate for a fine
mist,
this property of the scheme is incompatible
with the \citetalias{Fielding2022} framework, where
the current state of an individual cloud arises from the integration
of its past evolution, an essentially Lagrangian behaviour.

In another difference between our two implementations, \citet{Butsky2024} use a simplified form of 
the momentum transfer terms; compare our \cref{eq:ptrans}
to their equation 14. In their implementation, momentum
transfer follows the net mass flow. In the
\citetalias{Fielding2022} model clouds gain mass via
the TRML and lose mass through shredding simultaneously,
meaning that the sign of the net momentum transfer is not
constrained to have the same sign as the net mass transfer.
As can be seen in our \cref{eq:vdot}, the acceleration
of the cloud due to mass transfer always reduces
the relative velocity of the cloud and wind and its magnitude
only depends on the mass growth rate, not the mass loss rate.
This means that even if a cloud is losing mass overall (or its mass is not changing), it
is still accelerated by accreted wind material as long as $\dot{m}_\mathrm{grow} \neq 0$. This can be seen
in our \cref{fig:windtunnel,fig:ump_track}.
In the \citet{Butsky2024} scheme, only clouds that have a net mass growth
($\dot{m}=\dot{m}_\mathrm{grow} - \dot{m}_\mathrm{loss} > 0$)
can be accelerated by the accretion of hot wind material and, even then,
this acceleration is underestimated.
A related difference between our schemes arises
in the form given for the energy transfer from hot phase to
cool phase (their equation 18). Their prescription
allows energy transfer based on the magnitude
and sign of the net mass transfer, rather than considering
the mass growth and loss rates independently.
The net energy transfer should not
depend purely on the net mass transfer rate, but
on the growth and loss rates independently (see our \cref{eq:Edot}).
Additionally,
the energy transfer determined via their equation 18 enforces
the specific energy of the mass being transferred to always
be that of the hot phase, regardless of the sign of the net mass transfer.
The consequence is that as cool gas is mixed into the hot phase, the
specific energy of the resulting mixed gas is the same as the hot
phase before mixing. Thus, the cool gas has been
heated to the temperature of the hot phase, violating energy conservation.

\section{Conclusion} \label{sec:conclusion}
\textsc{Arkenstone} is a novel scheme implemented within the
\textsc{Arepo} code that allows the modelling
of multiphase stellar feedback-driven galactic winds within
coarse resolution cosmological hydrodynamic simulations
of galaxy formation. In
\citetalias{Smith2024}, we demonstrated aspects of the scheme
that allow high specific energy (i.e. hot and fast) winds
to be accurately modelled. This is particularly important
to properly capture the operation of such winds as
a form of preventative feedback. In this work,
we have presented the \textsc{Arkenstone}
cloud particle treatment, which permits the inclusion
of a population of cool clouds embedded in the hot, fast wind.
Resolving the relevant interactions between cool clouds
and the ambient medium in a cosmological galaxy formation
simulation is intractable because of the vast dynamic
range in spatial scales that must be captured.
However, their inclusion is necessary in order
to model a truly multiphase wind, while preceding
analytic work has demonstrated the
importance of their impact on high
specific energy outflows.

We model cool clouds using collisionless N-body particles
(which we term cloud particles) that can move
relative to the gas resolved by \textsc{Arepo}'s hydrodynamic scheme.
A cloud particle represents an unresolved population of
cool clouds. In addition to feeling gravity, the cloud particle
exchanges mass, momentum, energy, and metals with
the ambient gas with which it is co-located.
Models of cloud evolution and cloud--wind interactions, derived from high resolution
simulations of `cloud crushing' and turbulent radiative
mixing layers, can be distilled into the various
exchange terms, as well as into
the subgrid evolution of the cloud population.
\textsc{Arkenstone} is agnostic as to the
choice of cloud--wind interaction model,
granting substantial flexibility.
In addition to describing the underlying scheme,
in this work we presented an implementation
of the \citetalias{Fielding2022} cloud--wind
model into the \textsc{Arkenstone} framework.

We demonstrated the scheme using idealised non-cosmological
simulations carried out at a coarse resolution
appropriate for a large volume cosmological simulation.
The setup duplicated that used in \citetalias{Smith2024},
featuring a $M_{200}=10^{11}\ \Msun$ system
with a cooling flow CGM. We included a high
specific energy wind, then performed simulations
with varying input mass loadings for the cool
cloud phase, enabling us to illustrate
various characteristic regimes. The main
findings are as follows:

\begin{enumerate}
\item As expected from preceding analytic work,
the \citetalias{Fielding2022} cloud--wind
interaction model enables the efficient acceleration
of clouds embedded in the hot wind. All clouds
are accelerated, but their fate depends on their
initial mass. Initially low mass clouds are
rapidly accelerated, but are relatively
short lived. Higher mass clouds take longer
to be accelerated up to the background wind
velocity, but are more likely to survive.

\item The strong dependence of cloud
trajectories and mass evolution as a function
of initial cloud mass and current local wind
conditions gives rise to a complex distribution
of cloud masses throughout the wind.
Importantly, this means that the distribution
cannot be captured by instantaneous local
properties, making it impossible to capture with
a simpler statistical model. \label{item:cloud_distribution}

\item As shown in \citetalias{Smith2024}, the properties
of the background wind vary in both space and time.
There is a variation in the wind properties with polar angle,
the wind is collimated by the CGM (which changes the opening
angle of the wind with height) and changes of the SFR of the
galaxy over time result in fluctuations in wind properties.
Clouds take time to respond to changes in the local 
properties of the background wind, with less massive clouds
having a stronger coupling to the flow than more massive clouds.
This leads to a complex range of cloud trajectories, with
some clouds moving out of the wind and falling back towards
the galaxy. Some of these return to the ISM, but others
can be re-accelerated before they reach the galaxy. This
adds further complexity to the spatial distribution
of clouds in a way that cannot be captured in a simple
1D model.

\item The emergent mass outflow rate of cool clouds at some
distance from the galaxy is a non-monotonic function of the
input cool cloud mass loading factor because of the
back-reaction on the high specific energy wind.
When the input cool cloud mass loading is negligible,
the impact on the wind is similarly negligible, allowing
it to continue to efficiently regulate the galaxy's
SFR by suppressing CGM inflows (i.e. preventative feedback).
When we used a relatively high input cloud mass loading,
this lead to an initially high flux of cool clouds
and a noticeable drop in the SFR because of the increase
in the amount of mass being removed from the ISM (i.e. ejective feedback).
However, increasing the mass of clouds in the high
specific energy wind necessarily results in an increase in
the transfer of energy from the wind to the clouds.
This resulted in an eventual `poisoning' of the wind,
the resumption of strong inflows from the CGM
and a return to higher SFRs. We demonstrated
a simulation with an input cloud mass loading
between those two extremes. This was able
to fill a substantial portion of the halo
with cool clouds carried by the wind
without overly impacting the ability
of the hot wind to perform preventative
feedback. The exact boundaries of these
different regimes will depend strongly
on properties of the galaxy, CGM, halo, and mass
and energy loading of the high specific energy
wind.

\item Several phenomena illustrated in our simulations
indicate that two-fluid/multifluid schemes,
an alternative approach to that adopted by \textsc{Arkenstone},
are 
not as well suited for modelling cool clouds entrained in
a wind. While allowing a relative velocity between
the hot and cool phases, these schemes enforce
a single valued velocity field for the cool material.
We have demonstrated that on the scales
resolvable in this class of simulation the cool cloud
velocity field is multivalued because of the differential
acceleration of clouds with different masses and
the ubiquitous presence of intersecting
cloud trajectories (e.g. in fountain flows).
Additionally, as noted in \cref{item:cloud_distribution}, the
distribution of cloud properties at a given
location results from an integration of
their past trajectory and mass evolution in
a manner that cannot be derived from
instantaneous or local properties. It is
difficult to conceive of a method of
tracking this evolution with an Eulerian
approach as used in two-fluid/multifluid
schemes.
\end{enumerate}

In this work, we demonstrated \textsc{Arkenstone} with
a particular model for cloud--wind interactions
\citepalias{Fielding2022}, but the scheme is
agnostic to this choice. We can therefore use this
approach to study the large-scale consequences
of other theories of cloud evolution, which
may include additional physics such as
thermal conduction and magnetic fields.
We adopted constant mass and energy loading factors
to provide a clean numerical experiment,
but in future work these will be varied
according to halo, galaxy and/or ISM
properties.
We performed idealised non-cosmological
simulations in this work, but the
scheme has been specifically designed to
work in cosmological volume simulations;
Bennett et al. in prep. will demonstrate
some first cosmological applications of
\textsc{Arkenstone}.

\section*{Acknowledgements}
We are grateful to R{\"u}diger Pakmor for helpful comments.
This work was supported by the Simons Collaboration on “Learning the Universe.”
GLB acknowledges support from the NSF (AST-2108470, XSEDE grant MCA06N030), NASA TCAN award 80NSSC21K1053, and the Simons Foundation (grant 822237).
CGK and ECO acknowledge grant 10013948 from the Simons Foundation to Princeton University, to support the Learning the Universe Collaboration. 
Computations were performed on the HPC systems Raven and Freya at the Max Planck Computing and Data Facility (MPCDF).
The following open source software packages were used in this work:
\texttt{Astropy} \citep{AstropyCollaboration2013,AstropyCollaboration2018,AstropyCollaboration2022},
\texttt{Matplotlib} \citep{Hunter2007},
\texttt{nanoflann} \citep{Blanco2014},
\texttt{NumPy} \citep{Harris2020},
\texttt{SciPy} \citep{Virtanen2020}.

%%%%%%%%%%%%%%%%%%%%%%%%%%%%%%%%%%%%%%%%%%%%%%%%%%
\section*{Data Availability}

The data underlying this article will be shared on reasonable request to the corresponding author.



\bibliographystyle{mnras}
\bibliography{references}

%%%%%%%%%%%%%%%%%%%%%%%%%%%%%%%%%%%%%%%%%%%%%%%%%%

%%%%%%%%%%%%%%%%% APPENDICES %%%%%%%%%%%%%%%%%%%%%

%%%%%%%%%%%%%%%%%%%%%%%%%%%%%%%%%%%%%%%%%%%%%%%%%%

% Don't change these lines
\bsp	% typesetting comment
\label{lastpage}
\end{document}